\documentclass[aps,prd,groupedaddress,showpacs]{revtex4}
\usepackage{amsmath}
\usepackage[parfill]{parskip}    
\usepackage{float}
\usepackage{caption}
\usepackage{subcaption}
\usepackage{graphicx}
\usepackage{bbm}
\usepackage{graphicx}
\usepackage{amssymb}
\usepackage{epstopdf}
\usepackage{color}
\usepackage{ulem}
\usepackage{graphicx}
\usepackage{pdfpages}
\usepackage[colorlinks=true,citecolor=blue,urlcolor=magenta,breaklinks]{hyperref}

\begin{document}

\title{Geodesic analysis and black hole shadows on a general non--extremal rotating black hole in five--dimensional gauged supergravity }

\author{E. Contreras }
\email{econtreras@usfq.edu.ec}
\affiliation{Departamento de F\'isica, Colegio de Ciencias e Ingenier\'ia, Universidad San Francisco de Quito,  Quito, Ecuador\\}

\author{\'Angel Rinc\'on}
\email{angel.rincon@pucv.cl}
\affiliation{Instituto de F{\'i}sica, Pontificia Universidad Cat{\'o}lica de Valpara{\'i}so, Avenida Brasil 2950, Casilla 4059, Valpara{\'i}so, Chile\\}

\author{Grigoris Panotopoulos}
\email{grigorios.panotopoulos@tecnico.ulisboa.pt}
\affiliation{Centro de Astrof{\'i}sica e Gravita{\c c}{\~a}o, Departamento de F{\'i}sica, Instituto Superior T{\'e}cnico-IST, Universidade de Lisboa-UL, Av. Rovisco Pais, 1049-001 Lisboa, Portugal\\}

\author{Pedro \surname{Bargueño}}
\email{pedro.bargueno@ua.es}
\affiliation{Departamento de F\'{i}sica Aplicada, Universidad de Alicante, Campus de San Vicente del Raspeig, E-03690 Alicante, Spain}
\begin{abstract}
In this work,
motivated by the fact that
higher dimensional theories predict the existence of  black holes which differ from their four dimensional counterpart, 
we analyse the geodesics
and black hole shadow cast by a general non--extremal five dimensional black hole. The system under consideration corresponds to the 
Chong--Cveti\v{c}--L\"u--Pope
(Phys. Rev.Lett.{\bf 95}, 161301 (2005)), which has the Myers--Perry black hole as a limit.  
\end{abstract}

\maketitle

\section{Introduction}\label{intro}
Black holes (BHs) are a generic prediction of Einstein's General Relativity \cite{GR} (GR) and other metric theories of gravity, and they are believed to be formed in the gravitational collapse of massive stars during their final stages. 
BHs are the simplest objects in the Universe, characterized entirely by a handful of parameters, namely their mass, charges and rotation speed, although they are  more than mathematical objects.\\

The first BH solution was found by K. Schwarzschild
and describes  the
spacetime of a non-rotating BH \cite{SBH}. Later on, H. Reissner \cite{Re} and 
G. Nordström \cite{No} found a BH solution
describing
the spacetime of a non-rotating BH with a possible non-vanishing electrically
charge \cite{Bambi:2017khi}. 
The first rotating BH solution was found by R. Kerr in 1967 \cite{kerr} and further developments on this direction led to obtaining the charged rotating  
solution by E. Newman and co-workers in 1965 \cite{New}. As it is well known, the rotating BHs are more important in the astrophysical observations,
as the BH spin plays a critical and key role in any astrophysical process. In this respect, 
the development on rotating solutions has increase considerably in recent years (see 
\cite{Azreg-Ainou:2014nra,Azreg-Ainou:2014aqa,azreg2014,bobir17,quint1,azreg2019,Kim:2019hfp,Chen:2020aix,Contreras:2021yxe}, for example). Also, the interest in higher dimensional black hole physics is constantly increasing (see for instance \cite{Schee:2008kz,Stuchlik:2010zz,Stuchlik:2018qyz} and references therein.)
\\

From the experimental point of view, recent advances have led to the direct observations of BHs. In particular, on one hand, five years ago the LIGO collaboration 
directly detected for the first time the gravitational waves emitted from a BH merger of $\sim 60~M_{\odot}$ \cite{ligo}, but there was no information
on the defining property of BHs, which is no other than their horizon. On the other hand, around one year ago the Event Horizon Telescope (EHT)
project \cite{project}, observed a characteristic 
shadow-like image \cite{L1} (see also \cite{L2,L3,L4,L5,L6} for physical origin of the shadow, data processing and calibration, instrumentation etc), 
that is a darker region over a brighter background, via strong gravitational lensing and photon capture at the horizon. Thus, the observation of the 
shadow does probe the space-time geometry in the vicinity of the horizon, and doing so it tests the existence and properties of the latter \cite{psaltis}, 
although one should keep in mind that other horizon-less objects that possess light rings also cast shadows 
\cite{horizonless1,horizonless2,horizonless3,horizonless4,horizonless5,horizonless6,shakih2018,shakih2019},
and therefore the presence of a shadow does not necessarily implies that the object is a BH. Therefore, shadows as well as strong lensing \cite{vib1,vib2} 
images provide us with the exciting opportunity a) to detect the nature of a compact object, and b) to test whether the gravitational field around a
compact object is described by a rotating or non-rotating geometry. For a recent brief review on shadows see \cite{review}. The shadow of the Schwarzschild geometry was considered in \cite{synge,luminet}, while the shadow cast by the Kerr solution
was studied in \cite{bardeen} (see also \cite{monograph}). For shadows of Kerr BHs with scalar hair see \cite{carlos1,carlos2}, and for BH shadows in other frameworks see 
\cite{Bambi:2008jg,Bambi:2010hf,study1,study2,Moffat,quint2,study3,Schee:2016yzb,study4,study5,Schee:2017hof,study6,bobir2017,kumar2018,ovgun2018,Konoplya:2019sns,sudipta2019,shakih2019b,contreras2019rot,sabir19,sunnya,sunnyb,Konoplya:2019xmn,Konoplya:2019fpy,Allahyari:2019jqz,Tinchev:2019qwt,Cunha:2019ikd,Ovgun:2019jdo,Konoplya:2019goy,Hensh:2019ipu,Stuchlik:2019uvf,Contreras:2019cmf,Fathi:2019jid,Chang:2020lmg,Badia:2020pnh,Li:2020drn,Li:2019lsm,Khodadi:2020jij,Ovgun:2020gjz,Liu:2020ola,Vagnozzi:2020quf,Ghosh:2020ece}.
\\

In recent years the study of BHs in five dimensions and the shadow they cast \cite{38,39,40,41,42} has become in an active research line motivated by ideas in the braneworld, string theory and gauge/gravity duality \cite{Martin:2001te,Martin:2000xd,Martin:1999iy}. Kaluza--Klein theories \cite{1,2} and Superstring Theory \cite{3,4} suggest that extra spatial dimensions may exist. What is more, the remarkable AdS / CFT correspondence \cite{5,6} motivates the study of theories with a negative cosmological constant. In supergravity \cite{7}, which is the low-energy limit of superstring theory,  consistency of supersymmetry transformations requires a scalar potential for the scalar degrees of freedom, or the presence of a non-vanishing CC if there are no scalar degrees of freedom. Gauged supergravities \cite{8}, where the gravitinos are charged under the gauge fields, very often admit anti--de Sitter space as a vacuum state, and thus BH solutions of these theories are of physical relevance to the proposed AdS/CFT correspondence \cite{9}. Contrary to 4D SUGRA, where the bosonic part of the Lagrangian consists of the Einstein-Maxwell theory, and a single angular momentum for rotating solutions, in higher dimensions there are several angular momenta $J_i$ equal to the rank of the rotation group $SO(D-1)$ \cite{10}. Furthermore, in odd number of dimensions $D=5,7,...$ there is a also a Chern-Simons term, which although does not affect static (non-rotating)
solutions, does affect stationary (rotating) solutions \cite{11}. For charged, rotating BH solutions in 5D gauged SUGRA see e.g. \cite{39,9,11,12,13}  and references therein.  In the present work we propose to study
photon orbits via the Hamilton-Jacobi equations, and the shadow cast of the general, non--extremal  rotating BH solution in $5D$ gauged SUGRA with two--independent rotating parameters, obtained in \cite{39}.\\

In the present work, we will perform a geodesic analysis, and also we compute the black hole shadows of a  non-extremal rotating black hole in five-dimensional gauged supergravity. It should be pointed out that such analysis is made for the first time due to the complexity of the metric. In particular, although the method used here is not new, the implementation of that is quite complicated, and an analytical solution, although possible,  is highly non-trivial. We use the well-known celestial coordinates $(\alpha, \beta)$ to parametrize the black hole shadow.

Our work is organized as follows. In the next section, we briefly summarize the main aspects of the five dimensional solution in \cite{39}. In section \ref{HamJac}, we study the separability if the Hamilton--Jacobi equations while
in section \ref{ccoord} we deduce the suitable celestial coordinates for the study of the silhouette cast by our 5D-model and plot the associated shadows. Finally, we conclude our work in the last section.

\section{Five dimensional rotating BH solution}\label{five}
This section is devoted to review the main aspects on the Chong--Cveti\v{c}--L\"u--Pope (CCLP)
general non-extremal rotating BH in minimal five-dimensional gauged
supergravity reported in \cite{39}.\\

In terms of the Boyer--Lindquist like--coordinates, $(t,r,\theta,\phi,\psi)$,
the non--vanishing components of the metric are given by
\begin{eqnarray}
g_{00}&=&-\frac{\Delta_{\theta}(1+g^{2}r^{2})}{\Xi_{a}\Xi_{b}}+\frac{\Delta_{\theta}^{2}(2m\rho^{2}-q^{2}+2abqg^{2}\rho^{2})}{\rho^{4}\Xi_{a}^{2}\Xi_{b}^{2}}\nonumber\\
g_{03}&=&-\frac{\Delta_{\theta}[a(2m\rho^{2}-q^{2})+bq\rho^{2}(1+a^{2}g^{2})]\sin^{2}\theta}{\rho^{4}\Xi_{a}^{2}\Xi_{b}}\nonumber\\
g_{04}&=&-\frac{\Delta_{\theta}[b(2m\rho^{2}-q^{2})+aq\rho^{2}(1+b^{2}g^{2})]\cos^{2}\theta}{\rho^{4}\Xi_{b}^{2}\Xi_{a}}\nonumber\\
g_{33}&=&\frac{(r^{2}+a^{2})\sin^{2}\theta}{\Xi_{a}}+\frac{a[a(2m\rho^{2}-q^{2})+2bq\rho^{2}]\sin^{4}\theta}{\rho^{4}\Xi_{a}^{2}}\nonumber\\
g_{44}&=&\frac{(r^{2}+b^{2})\cos^{2}\theta}{\Xi_{b}}+\frac{b[b(2m\rho^{2}-q^{2})+2aq\rho^{2}]\cos^{4}\theta}{\rho^{4}\Xi_{a}^{2}}\nonumber\\
g_{34}&=&\frac{[ab(2m\rho^{2}-q^{2})+(a^{2}+b^{2})q\rho^{2}]\sin^{2}\theta \cos^{2}\theta}{\rho^{4}\Xi_{a}\Xi_{b}}\nonumber\\
g_{11}&=&\frac{\rho^{2}}{\Delta_{r}}\nonumber\\
g_{22}&=&\frac{\rho^{2}}{\Delta_{\theta}},
\end{eqnarray}
where
\begin{eqnarray}
\Delta_{\theta}&=&1-a^{2}g^{2}\cos^{2}\theta-b^{2}g^{2}\sin^{2}\theta\\
\Delta_{r}&=&\frac{(r^{2}+a^{2})(r^{2}+b^{2})(1+g^{2}r^{2})+q^{2}+2abq}{r^{2}}-2m\\
\rho^{2}&=&r^{2}+a^{2}\cos^{2}\theta+b^{2}\sin^{2}\theta\\
\Xi_{a}&=&1-a^{2}g^{2}\nonumber\\
\Xi_{b}&=&1-b^{2}g^{2},
\end{eqnarray}
which solve the equations of motion of minimal gauged five-dimensional supergravity, which follow from the Lagrangian \cite{39}
\begin{eqnarray}
\mathcal{L}=(R+12g^{2})*\mathbbm{1}
-\frac{1}{2}*F\wedge F+\frac{1}{3\sqrt{3}}F\wedge F\wedge A,
\end{eqnarray}
with $F=dA$ and $g>0$. Note that, the other well known five dimensional Myers--Perry BH solution is recovered in the limits $g\to0$ and $q\to0$, namely
\begin{eqnarray}
g_{00}^{MP}&=&-1+\frac{2 m}{\rho^{2}}\\
g_{03}^{MP}&=&-\frac{2 a m \sin ^2(\theta )}{\rho^{2}}\\
g_{04}^{MP}&=&-\frac{2 b m \cos ^2(\theta )}{\rho^{2}}\\
g_{33}^{MP}&=&\sin ^2\theta  \left(\frac{2 a^2 m \sin ^2\theta }{\rho^{2}}+a^2+r^2\right)\\
g_{44}^{MP}&=&\cos ^2\theta  \left(\frac{2 b^2 m \cos ^2\theta }{\rho^{2}}+b^2+r^2\right)\\
g_{34}^{MP}&=&\frac{2 a b m \sin ^2\theta  \cos ^2\theta }{\rho^{2}}\\
g_{11}^{MP}&=&\frac{r^{2}\rho^{2}}{r^2 \left(a^2+b^2-2 m\right)+a^2 b^2+r^4}\\
g_{22}^{MP}&=&\rho^{2}.
\end{eqnarray}

A particularity of CCLP model is that, under certain considerations, we can
ensure both the BPS limit and the avoidance of naked closed time-like curves (CTC's) of the solution. To this end
the set of parameters $\{q,a,b,g,m\}$ must be restricted as 
\begin{eqnarray}
q&=&\frac{m}{1+(a+b)g}\label{q}\\
gm&=&(a+b)(1+ag)(1+bg)(1+ag+bg)\label{gm},
\end{eqnarray}
form where the function $\Delta_{r}$ has a double root and now is given by
\begin{eqnarray}
\Delta_{r}=r^{-2}(r^{2}-r_{0}^{2})^{2}[g^{2}r^{2}+(1+ag+bg)],
\end{eqnarray}
with $r_{0}^{2}=g^{-1}(a+b+abg)$. Note that, at the Killing horizon $r=r_{0}$, the determinant of the metric in the $(\theta,\phi,\psi)$ direction reads
\begin{eqnarray}
\frac{(a+b)^{2}(a+b+abg)\sin^{2}\theta}{4g^{3}(1-ag)^{2}(1-bg)^{2}},
\end{eqnarray}
which implies that the remaining free parameters must be restricted to
\begin{eqnarray}
a+b+abg>0,
\end{eqnarray}
in order the no apparition of CTC's is ensured. It is worth noticing that restriction (\ref{q}) and (\ref{gm}) characterized the extremal solution.\\

\begin{figure}[h!]
\centering
\includegraphics[scale=0.5]{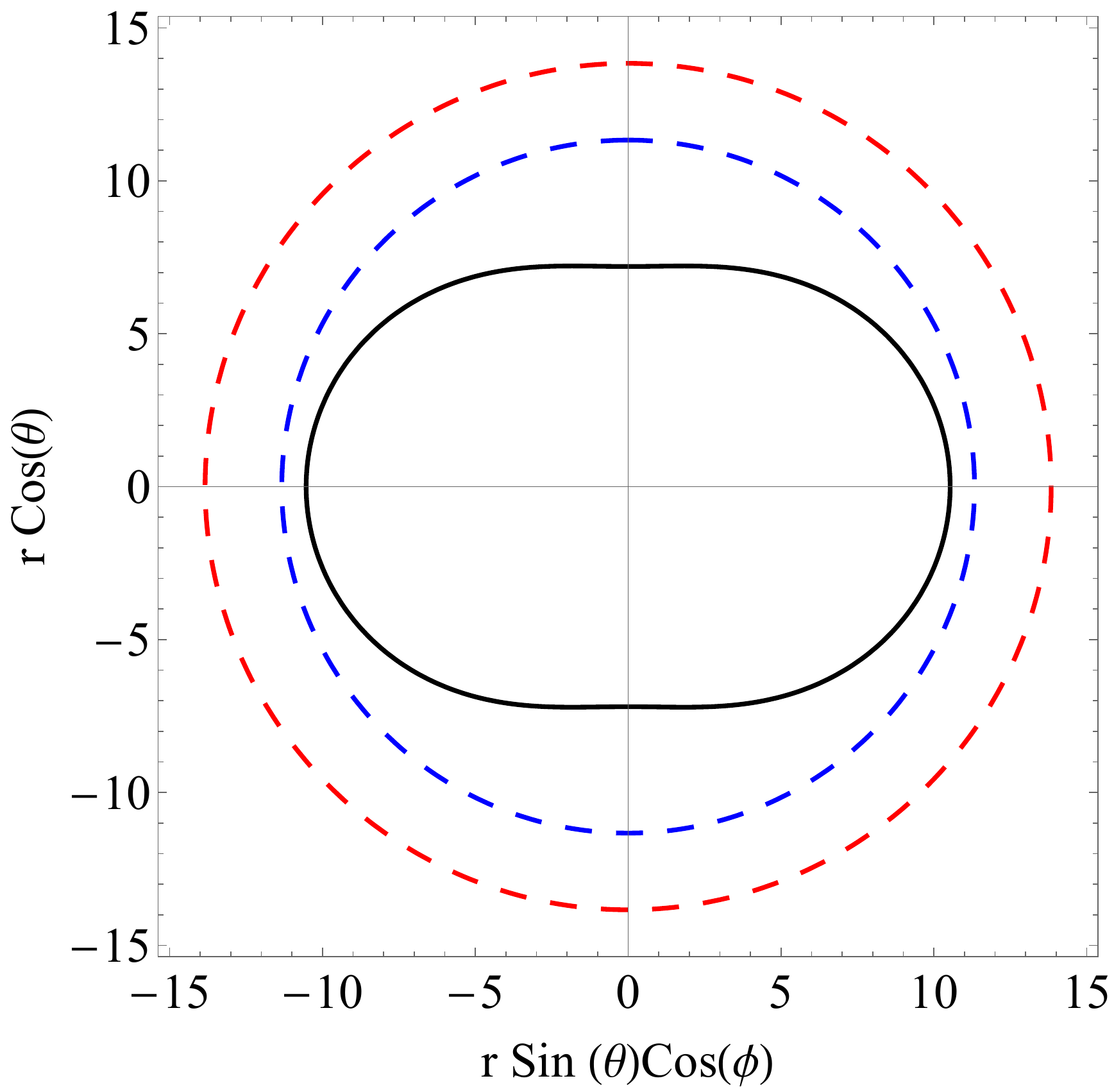}
\caption{\label{ergo} 
Ergoregion for the extremal CCLP (black solid line), the non-extremal CCLP case (blue dashed line) and MP BH (red dashed line). All the figures have been plotted assuming
$a=4$, $b=1$ and $m=100$. For the non--extremal CCLP case we have assume $g=0.1$,  $q=80$.
}
\end{figure}

Another interesting feature to be discussed corresponds to the static limit surface, namely, the region where any observer cannot remain at rest and cannot be static which requires $g_{00}=0$. The
behaviour of the outer ergoregion in $x- z$ plane is depicted in fig. \ref{ergo} for both, the CCLP and MP BH solutions. 
It should be mentioned that although the CCLP and the MP solutions depend on a different set of parameters, in Fig. \ref{ergo} we have used the same values for the parameters they share, namely $\{a,b,m\}$, with the aim to compare the results. In this regard, the ergoregion defined by the MP model is greater than both the non--extremal and extremal shown by the CCLP solution.

\section{Hamilton--Jacobi equations and BH shadows}\label{HamJac}
The Hamilton-Jacobi formalism represents an efficient method to derive the equations of motion for test particles because instead of solving second order differential equations arising from the geodesic equation, it leads to first order equations,
which significantly simplifies the calculations \cite{Bambi:2017khi}. In this section we shall follow the Hamilton-Jacobi formalism to obtain the geodesic equations for the CCLP model. It should be emphasized that, to the best of our knowledge, this is the first time that the geodesic equations for the CCLP solution is found the manner we show here.\\

Let us start with the Hamilton--Jacobi equations written in the form
\begin{eqnarray}\label{HJ}
2\frac{\partial S}{\partial \tau}=g^{\mu\nu}\frac{\partial S}{\partial x^{\mu}}\frac{\partial S}{\partial x^{\nu}},
\end{eqnarray}
where $S$ is the Jacobi action and $\tau$ is the proper time. By assuming the separability of the Jacobi action, we propose
\begin{eqnarray}\label{action}
S=\frac{\epsilon}{2}\tau-E t+L_{\phi}\phi+L_{\psi}\psi+S_{r}(r)+S_{\theta}(\theta)
\end{eqnarray}
where $\epsilon=0$ for null geodesics and $\epsilon=1$ for time-like ones. Furthermore, $E, L_{\phi}$ 
and $L_{\psi}$ are the conserve energy and momenta per unit mass. Replacing (\ref{action})
in (\ref{HJ}) we arrive at
\begin{eqnarray}
\epsilon&=&g^{00}E^{2}-2g^{03}E L_{\phi}
-2g^{04}EL_{\psi}+g^{33}L_{\phi}^{2}+g^{44}L_{\psi}^{2}\nonumber\\
&&+2g^{34}L_{\phi}L_{\psi}+g^{11}\left(\frac{\partial S_{r}}{\partial r}\right)^{2}+
g^{22}\left(\frac{\partial S_{\theta}}{\partial \theta}\right)^{2}\label{hje}.
\end{eqnarray}
For the CCLP solution, $g^{\mu\nu}$ components of the metric are given by
\begin{eqnarray}
g^{00}&=&\frac{1}{\rho^{2}}\left(
\frac{\mathcal{A}(r)}{\Delta_{r}}-
\frac{1}{g^{2}}\left(\frac{1}{\Delta_{\theta}}-\frac{1}{1+g^{2}r^{2}}\right)
\Xi_{a}\Xi_{b}\right)\label{gu00}\\
g^{11}&=&\frac{\Delta_{r}}{\rho^{2}}\label{gu11}\\
g^{22}&=&\frac{\Delta_{\theta}}{\rho^{2}}\label{gu22}\\
g^{33}&=&\frac{1}{\rho^{2}}\left(
\frac{\mathcal{B}(r)}{\Delta_{r}}+
\left(
\frac{1}{\sin^{2}\theta}
+\frac{b^{2}-a^{2}}{a^{2}+r^{2}}
\right)\Xi_{a}
\right)\label{gu33}\\
g^{44}&=&\frac{1}{\rho^{2}}\left(
\frac{\mathcal{C}(r)}{\Delta_{r}}+
\left(
\frac{1}{\cos^{2}\theta}
+\frac{a^{2}-b^{2}}{b^{2}+r^{2}}
\right)\Xi_{b}
\right)\label{gu44}\\
g^{03}&=&\frac{\mathcal{D}(r)}{\rho^{2}\Delta_{r}}\label{gu03}\\
g^{04}&=&\frac{\mathcal{E}(r)}{\rho^{2}\Delta_{r}}\label{gu04}\\
g^{34}&=&\frac{\mathcal{F}(r)}{\rho^{2}\Delta_{r}}\label{gu34}
\end{eqnarray}
where, after a long computation,  $\mathcal{A},\mathcal{B}, \mathcal{C},\mathcal{D},\mathcal{E}$ and 
$\mathcal{F}$ read
\begin{widetext}
\begin{eqnarray}
\mathcal{A}&=&-\frac{a b g^2 q \left(2 a^2 b^2+2 a^2 r^2+a b q+2 b^2 r^2+2 r^4\right)
-\left(a^2+b^2+r^2\right) \left(q^2-2 m r^2\right)
+2 a^2 b^2 m}{\left(g^2 r^2+1\right)}\\
\mathcal{B}&=&\frac{a^2 \left(-b^2 \left(2 g^2 m r^2-g^2 q^2+2 m\right)+g^2 r^2 \left(q^2-2 m r^2\right)-2 m r^2+q^2\right)-2 a b q \left(b^2+r^2\right) \left(g^2 r^2+1\right)-b^2 q^2}{ \left(a^2+r^2\right)}\\
\mathcal{C}&=&
\frac{b^2 \left(-a^2 \left(2 g^2 m r^2-g^2 q^2+2 m\right)+g^2 r^2 \left(q^2-2 m r^2\right)-2 m r^2+q^2\right)-2 a b q \left(a^2+r^2\right) \left(g^2 r^2+1\right)-a^2 q^2}{ \left(b^2+r^2\right) }\\
\mathcal{D}&=&-(a^2 \left(b^3 g^2 q+b g^2 q r^2\right)+a \left(2 b^2 m+2 m r^2-q^2\right)+b^3 q+b q r^2)\\
\mathcal{E}&=&-
(b^2 \left(a^3 g^2 q+a g^2 q r^2\right)+b \left(2 a^2 m+2 m r^2-q^2\right)+a^3 q+a q r^2)\\
\mathcal{F}&=&-(a^2 q \left(g^2 r^2+1\right)+a b \left(2 g^2 m r^2-g^2 q^2+2 m\right)+b^2 q \left(g^2 r^2+1\right))
\end{eqnarray}
\end{widetext}
Replacing (\ref{gu00}), (\ref{gu11}), (\ref{gu22}), (\ref{gu33}), (\ref{gu44}), (\ref{gu03}), (\ref{gu04}), (\ref{gu34}) in (\ref{hje}) we obtain
\begin{widetext}
\begin{eqnarray}
&&\Delta_{r}\left(\frac{\partial S_{r}}{\partial r}\right)^{2}+
\left(
\mathcal{A}+\frac{\Xi_{a}\Xi_{b}\Delta_{r}}{g^{2}(1+g^{2}r^{2})}\right)\frac{E^{2}}{\Delta_{r}}
+\left(\mathcal{B}+\frac{\Xi_{a}(b^{2}-a^{2})\Delta_{r}}{r^{2}+a^{2}}\right)\frac{L_{\phi}^{2}}{\Delta_{r}}+
\left(\mathcal{C}+\frac{\Xi_{b}(a^{2}-b^{2})\Delta_{r}}{r^{2}+b^{2}}\right)\frac{L_{\psi}^{2}}{\Delta_{r}}\nonumber\\
&&\hspace{8cm}+2\frac{\mathcal{F}L_{\phi}L_{\psi}}{\Delta_{r}}-2(\mathcal{D}L_{\phi}+\mathcal{E}L_{\psi})\frac{E}{\Delta_{r}}-\epsilon r^{2}=-\mathcal{Q}\label{dsr}\\
&&\Delta_{\theta}\left(\frac{\partial S_{\theta}}{\partial\theta}\right)^{2}-\frac{\Xi_{a}\Xi_{b}E^{2}}{g^{2}\Delta_{\theta}}+\Xi_{a}\csc^{2}\theta L_{\phi}^{2}+\Xi_{b}\sec^{2}\theta L_{\psi}^{2}
-\frac{\epsilon}{g^{2}}(1-\Delta_{\theta})=\mathcal{Q}\label{dsteta}
\end{eqnarray}
\end{widetext}
with $\mathcal{Q}$ the Carter constant \cite{carter}. Next, Eqs. (\ref{dsr}) and (\ref{dsteta})
can be formally integrated to yield
\begin{eqnarray}
S_{r}&=&\int^{r}\frac{\sqrt{R}}{\Delta_{r}}dr\\
S_{\theta}&=&\int^{\theta}\frac{\sqrt{\Theta}}{\sqrt{\Delta_{\theta}}}d\theta,
\end{eqnarray}
where
\begin{widetext}
\begin{eqnarray}
R&=&-\mathcal{Q}\Delta_{r}
-\left(
\mathcal{A}+\frac{\Xi_{a}\Xi_{b}\Delta_{r}}{g^{2}(1+g^{2}r^{2})}\right)E^{2}-2\mathcal{F}L_{\phi}L_{\psi}
-\left(\mathcal{B}+\frac{\Xi_{a}(b^{2}-a^{2})\Delta_{r}}{r^{2}+a^{2}}\right)L_{\phi}^{2}-
\left(\mathcal{C}+\frac{\Xi_{b}(a^{2}-b^{2})\Delta_{r}}{r^{2}+b^{2}}\right)L_{\psi}^{2}\nonumber\\
&&\hspace{10cm}+2(\mathcal{D}L_{\phi}+\mathcal{E}L_{\psi})E+\epsilon \Delta_{r}r^{2}\\
\Theta&=&\mathcal{Q}+\frac{\Xi_{a}\Xi_{b}E^{2}}{g^{2}\Delta_{\theta}}-\Xi_{a}\csc^{2}\theta L_{\phi}^{2}-\Xi_{b}\sec^{2}\theta L_{\psi}^{2}
+\frac{\epsilon}{g^{2}}(1-\Delta_{\theta}).
\end{eqnarray}
\end{widetext}
The Jacobi action reads now
\begin{eqnarray}
S=\frac{\epsilon\tau}{2}-E t+L_{\phi}\phi+L_{\psi}\psi+\nonumber\\
\int^{r}\frac{\sqrt{R}}{\Delta_{r}}dr+\int^{\theta}\frac{\sqrt{\Theta}}{\sqrt{\Delta_{\theta}}}d\theta.
\end{eqnarray}
Finally, the complete geodesic equations of the CCLP BH are obtained by taking variations of the Jacobi action respect to the parameters $\{\epsilon, E,L_{\phi},L_{\psi}\}$,
\begin{widetext}
\begin{eqnarray}
\rho^{2}\dot{t}&=&\frac{1}{\Delta_{r}}\left(\mathcal{A}+\frac{\Xi_{a}\Xi_{b}\Delta_{r}}{g^{2}(1+g^{2}r^{2})}
-\frac{\mathcal{D}L_{\phi}+\mathcal{E}L_{\psi}}{E}
\right)
E
-\frac{\Xi_{a}\Xi_{b}}{g^{2}\Delta_{\theta}}\label{tpunto}\\
\rho^{2}\dot{\phi}&=&-
\frac{1}{\Delta_{r}}
\left(
\mathcal{F}L_{\psi}+\left(\mathcal{B}+\frac{\Xi_{a}(b^{2}-a^{2})\Delta_{r}}{r^{2}+a^{2}}\right)L_{\phi}
-\mathcal{D}E\right)-\Xi_{a}\csc^{2}\theta L_{\phi}\label{fipunto}\\
\rho^{2}\dot{\psi}&=&
-\frac{1}{\Delta_{r}}\left(
\mathcal{F}L_{\phi}+\left(
\mathcal{C}+\frac{\Xi_{b}(a^{2}-b^{2})\Delta_{r}}{r^{2}+b^{2}}
\right)L_{\psi}-\mathcal{E}E\right)
-\Xi_{b}\sec^{2}\theta L_{\psi}\label{psipunto}\\
\rho^{4}\dot{r}^{2}&=&R\label{rad}\\
\rho^{4}\dot{\theta}^{2}&=&\Delta_{\theta}\Theta\label{theta},
\end{eqnarray}
\end{widetext}
where the dot represents derivative with respect
to $\tau$. We remark that this is the first time that Eqs. (\ref{tpunto}), (\ref{fipunto}), (\ref{psipunto}), (\ref{rad}) and (\ref{theta}) are presented accordance to the like--Boyer--Lindquist coordinates.\\

\begin{figure*}[hbt!]
\centering
\includegraphics[width=0.48\textwidth]{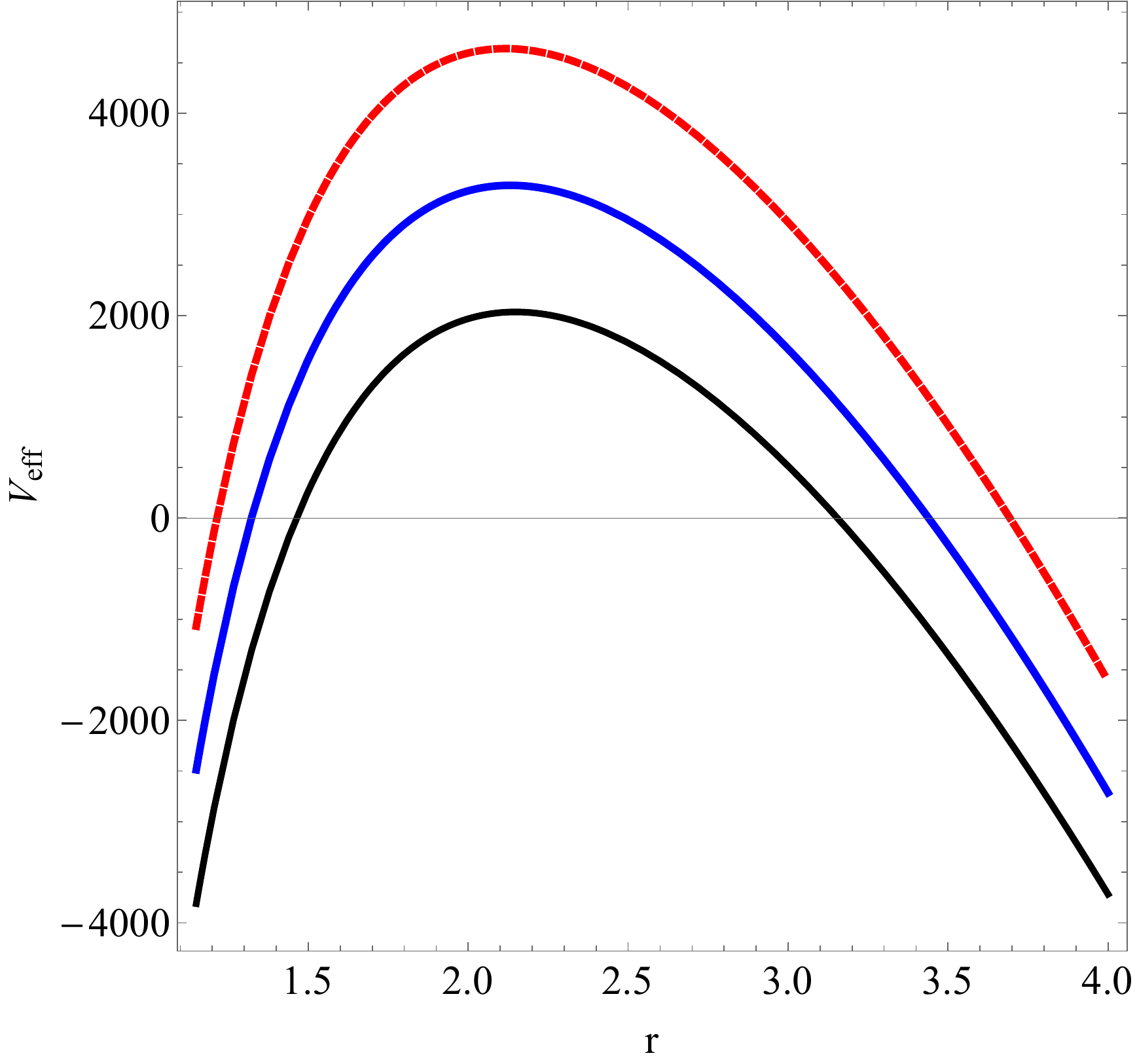}\
\includegraphics[width=0.48\textwidth]{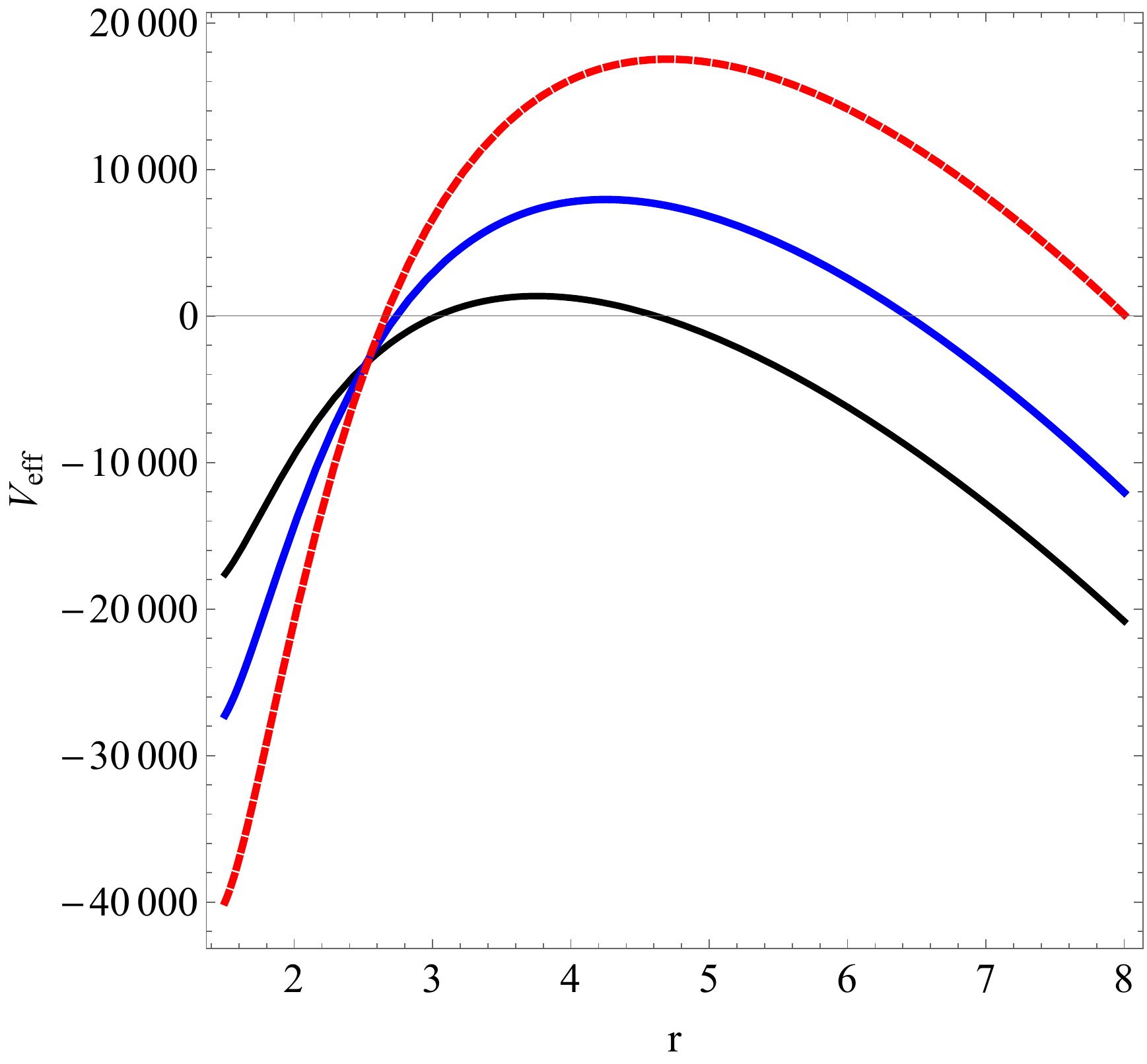}
\caption{\label{veff} Effective potential, $V_{eff}$ as a function of $r$ for $a=0.4$, $m=0.1$, $g=0.05$, $\mathcal{Q}/E^{2}=1$ and $L_{\psi}=0$. In left panel we set $b=a$ and $L_{\phi}/E=60$ (black), $L_{\phi}/E=60$ (blue) and $L_{\phi}/E=70$ (red). In the right panel $b=0.1$ and $L_{\phi}/E=150$ (black), $L_{\phi}/E=200$ (blue) and $L_{\phi}/E=250$ (red) }
\end{figure*}

For $\epsilon=0$, the above equations describe the propagation of light around our five dimensional BH. In this respect, the information about the silhouette of the BH in encoded in the radial motion by Eq. (\ref{rad}) with $\epsilon=0$
from where the effective potential reads
\begin{eqnarray}
V_{eff}&=&-\mathcal{Q}\Delta_{r}
-\left(
\mathcal{A}+\frac{\Xi_{a}\Xi_{b}\Delta_{r}}{g^{2}(1+g^{2}r^{2})}\right)E^{2}-2\mathcal{F}L_{\phi}L_{\psi}
-\left(\mathcal{B}+\frac{\Xi_{a}(b^{2}-a^{2})\Delta_{r}}{r^{2}+a^{2}}\right)L_{\phi}^{2}-
\left(\mathcal{C}+\frac{\Xi_{b}(a^{2}-b^{2})\Delta_{r}}{r^{2}+b^{2}}\right)L_{\psi}^{2}\nonumber\\
&&\hspace{10cm}+2(\mathcal{D}L_{\phi}+\mathcal{E}L_{\psi})E.
\end{eqnarray}
As it is well known, the effective potential for the photon can be though as a barrier with a maximum which comes from minus infinity (behind the horizon) and to a constant value
for $r\to\infty$ (see figure \ref{veff}).
Even more, the path taken by the photons could be of three types depending of their energy, $E$, as they approach to the maximum of $V_{eff}$. The first case corresponds to photons that arrive from infinity with $E>V^{MAX}_{eff}$, so that they are scattered by the BH, penetrate the horizon and fall down into the black hole. The second case corresponds to these photons arriving to the BH with $E<V^{MAX}_{eff}$, so that they are scattered around the BH and come back to infinity. The third case corresponds to photons with energy $E=V_{eff}^{MAX}$, so that this is a unstable orbit with constant radius. Is precisely the unstable orbits which are the responsible of the shadow observed, namely, observers located far from the BH associate a silhouette with the information of the photons with energy $E\gtrapprox V^{MAX}_{eff}$. Photons fulfilling that condition can barely escape from falling into the black hole and focus in a celestial plane.\\

From the above discussion we can define the condition for unstable orbits as $V_{eff}=0$ and $\partial V_{eff}/\partial r=0$. This condition leads to expression involving the impact parameters $\eta=\mathcal{Q}/E^{2}$, $\xi_{1}=L_{\phi}/E$ and $\xi_{2}=L_{\psi}/E$ which, given the nature of the solution, are 
too long to be shown here. However, in some particular cases the expressions can be simplified as we shall discuss in what follows.\\

First, let us define the Lagrangian
\begin{eqnarray}
L=\frac{1}{2}g_{\mu\nu}\dot{x}^{\mu}\dot{x}^{\nu},
\end{eqnarray}
from where, given that $\{t,\phi,\psi\}$ are cyclic, it is straightforward to obtain the conserved quantities $\{E,L_{\psi},L_{\phi}\}$ as follows
\begin{eqnarray}
-E&=&\frac{\partial L}{\partial \dot{t}}=
g_{00} \dot{t}+g_{03}\dot{\phi }+g_{04} \dot{\psi }\\
L_{\phi}&=&\frac{\partial L}{\partial \dot{\phi}}=
g_{03} \dot{t}+g_{33} \dot{\phi }+g_{34} \dot{\psi }\\
L_{\psi}&=&\frac{\partial L}{\partial\dot{\psi}}=
g_{04} \dot{t}+g_{34} \dot{\phi }+g_{44} \dot{\psi }\label{lpsi}.
\end{eqnarray}
Note that, for $\theta=0$, $L_{\phi}=0$ and for $\theta=\frac{\pi}{2}$, $L_{\psi}=0$. Without lost of generality, let us consider $\xi_{2}=0$ ($\theta=\pi/2$) so the conditions
$V_{eff}=0$ and $\partial V_{eff}/\partial r=0$ leads to 
\begin{eqnarray}
\mathfrak{a}\xi_{1}^{2}+\mathfrak{b}\xi_{1}+\mathfrak{c}=0
\end{eqnarray}
with
\begin{eqnarray}
\mathfrak{a}&=&\mathcal{B}\Delta_{r}'-\mathcal{B}'\Delta_{r}+\frac{2r(b^{2}-a^{2})\Xi_{a}\Delta_{r}^{2}}{(a^{2}+r^{2})^{2}}\\
\mathfrak{b}&=&2(\mathcal{D}'\Delta_{r}-\mathcal{D}\Delta_{r}')\\
\mathfrak{c}&=&\mathcal{A}\Delta_{r}'-\mathcal{A}'\Delta_{r}+\frac{2r\Xi_{a}\Xi_{b}\Delta_{r}^{2}}{(1+g^{2}r^{2})^{2}}.
\end{eqnarray}
For $\eta$, the solution can be written as
\begin{eqnarray}
\eta=-\bigg(
\frac{\mathcal{B}}{\Delta_{r}}+\frac{\Xi_{a}(b^{2}-a^{2})}{r^{2}+a^{2}}
\bigg)\xi_{1}^{2}+\frac{2\mathcal{D}}{\Delta_{r}}\xi_{1}-\frac{\mathcal{A}}{\Delta_{r}}-\frac{\Xi_{a}\Xi_{b}}{g^{2}(1+g^{2}r^{2})}.
\end{eqnarray}

Of course, for $\theta=0$ ($\xi_{1}=0$) the expressions are formally the same so we will not consider this case separately here.

\section{Celestial coordinates and shadows}\label{ccoord}
\begin{figure*}[h!]
\centering
\includegraphics[width=\textwidth]{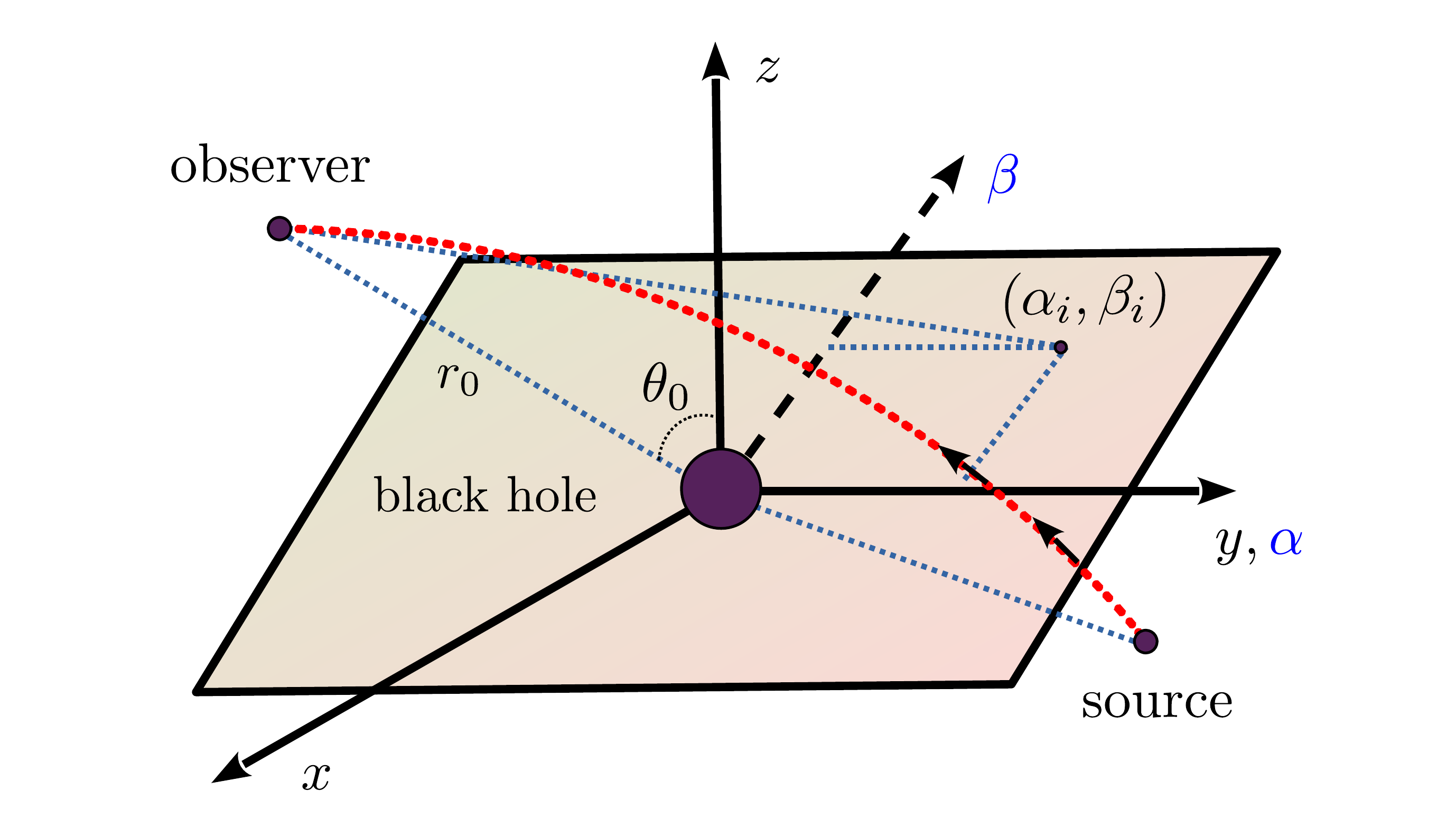} 
\caption{\label{shadow}
The figure shows a pictorial representation of how the light emitted from specific sources is curved by a black hole up to be received by the observer finally. Also, the corresponding celestial coordinates are included to clarify its geometrical meaning. 
 }
\end{figure*}

Let us consider an observer far form the black hole at an distance $r_{0}$ and with certain inclination angle $\theta_{0}$ between the line of the sight of the observer and the axis of rotation the BH. For this observer, the shadow of the black hole is projected in a celestial plane which axis we label as $(\alpha,\beta)$ and that, following \cite{goshEPJC}, can be written as 
\begin{eqnarray}
\alpha&=&\lim\limits_{r_{0}\to\infty}-\frac{
r_{0}}{\vartheta +\varphi \xi_{1}+\varrho \xi_{2}}\left(\frac{\xi_{1}}{\sqrt{g_{33}}}
+\frac{\xi_{2}}{\sqrt{g_{44}}}\right)\\
\beta&=&\pm\lim\limits_{r_{0}\to\infty}
\frac{
r_{0}}{\vartheta +\varphi \xi_{1}+\varrho \xi_{2}}\frac{\sqrt{\Theta}}{\sqrt{g_{22}}E},
\end{eqnarray}
with $\vartheta=\sqrt{-g^{00}}$, 
$\varphi=\sqrt{-g^{03}}$ and $\varrho=\sqrt{-g^{04}}$. 
Be aware and notice that, similarly to the four-dimensional case, this case can also be well understood by the above-mentioned standard celestial coordinates $(\alpha, \beta)$. Thus, such coordinates define a plane, and the silhouette is embedded in such a plane. What is more, the corresponding shadow is then obtained whether we take into account equatorial orbits of photon around the 5-dimensional black hole from the effective potential. For a pictorial representation of the physical situation, see Fig.~\eqref{shadow}.

However, it is worth noticing that, as the solution is not asymptotically flat but anti--de Sitter, the limit process should not to be taken to $r_{0}\to\infty$ but to some finite value. More precisely, as stated in \cite{76}
in the context of shadows of Kerr-Newman-NUT black
holes with a cosmological constant, the anti--de Sitter case the
domain of outer communication of the BH, namely,
the region between $r\to\infty$ and the first horizon corresponds
to the region where we can place observers for observing
the shadow of the BH.

In the present case, we have
\begin{eqnarray}
\vartheta&\approx&\sqrt{\frac{\Xi_{a}\Xi_{b}}{\Delta_{\theta}g^{2}r_{0}^{2}}}\\
\varphi&=&\varrho\approx0\\
g_{22}&\approx&\frac{r_{0}^{2}}{\Delta_{\theta}}\\
g_{33}&\approx&\frac{r_{0}^{2}}{\Xi_{a}}\sin^{2}\theta\\
g_{44}&\approx&\frac{r_{0}^{2}}{\Xi_{b}}\cos^{2}\theta,
\end{eqnarray}
form where the celestial 
coordinates take the form
\begin{eqnarray}
\tilde{\alpha}&=&\frac{\xi_{1}\csc\theta}{\sqrt{\frac{\Xi_{b}}{\Delta_{\theta}g^{2}}}}+\frac{\xi_{2}\sec\theta}{\sqrt{\frac{\Xi_{a}}{\Delta_{\theta}g^{2}}}}\label{alfaMP}\\
\tilde{\beta}&=&\frac{g\Delta_{\theta}}{\sqrt{\Xi_{a}\Xi_{b}}}\sqrt{\eta+\frac{\Xi_{a}\Xi_{b}}{g^{2}\Delta_{\theta}}
-\Xi_{a}\csc^{2}\theta\xi_{1}^{2}
-\Xi_{b}\sec^{2}\theta \xi_{2}^{2}}\label{betaMP},
\end{eqnarray}
with $\tilde{\alpha}=-\alpha/r_{0}$ and $\tilde{\beta}=\beta/r_{0}$ the normalized celestial coordinates. It should be highlighted that, to the best of our knowledge, this is the first time that celestial coordinates for the CCLP model have been express as Eqs. (\ref{alfaMP}) and (\ref{betaMP}). Besides, it is worth mentioning that results (\ref{alfaMP}) and (\ref{betaMP}) differ from the celestial coordinates usually reported in the literature. The main reason is that, in our case, the metric depends on $\Xi_{a},\Xi_{b},\Delta_{\theta}$ which appear explicitly in our expressions. Of course, if we discard one of the angular momentum parameters or consider an extra constraint between the parameters of the solution as usually done in the literature, we recover the standard expression for the celestial coordinates. More precisely, if we follow the same procedure with the MP solution, the expressions for $\alpha$ and $\beta$ correspond to those previously reported as we demonstrate in what follows.\\

In this case, the inverse components of the metric are given by 
\begin{eqnarray}
g^{00}_{MP}&=&-1-\frac{2 m \left(a^2+r^2\right) \left(b^2+r^2\right)}{\Delta \rho^2}\\
g^{11}_{MP}&=&\frac{\Delta }{r^2 \rho^2}\\
g^{22}_{MP}&=&\frac{1}{\rho^{2}}\\
g^{33}_{MP}&=&\frac{b^4-a^2 b^2-2 b^2 m-a^2 r^2+b^2 r^2}{\Delta  \rho ^2}+\frac{\csc ^2\theta }{\rho ^2}\\
g^{44}_{MP}&=&\frac{a^4-a^2 b^2-2 a^2 m+a^2 r^2-b^2 r^2}{\Delta  \rho ^2}+\frac{ \sec ^2\theta }{ \rho ^2}\\
g^{03}_{MP}&=&-\frac{2 a m \left(b^2+r^2\right)}{\Delta  \rho ^2}\\
g^{04}_{MP}&=&-\frac{2 b m \left(a^2+r^2\right)}{\Delta  \rho ^2}\\
g^{34}_{MP}&=&-\frac{2 a b m}{\Delta  \rho ^2}\\
\Delta&=&(r^{2}+a^{2})(r^{2}+b^{2})-2mr^{2}\\
\rho^{2}&=&r^{2}+a^{2}\cos^{2}\theta+b^{2}\sin^{2}\theta,
\end{eqnarray}
which can be easily obtained after taking the limits $q\to0$ and $g\to0$ in
Eqs. (\ref{gu00}),({\ref{gu11}}), (\ref{gu22}), (\ref{gu33}), (\ref{gu03}), (\ref{gu04}), (\ref{gu34}). Now, it is straightforward to show that, for the MP case, the Hamilton--Jacobi equations can be separated as
\begin{eqnarray}
\frac{1}{r}\frac{\partial \tilde{S}_{r}}{\partial r}=\frac{\sqrt{\tilde{R}}}{\Delta}\\
\frac{\partial \tilde{S}_{\theta}}{\partial \theta}=\sqrt{\tilde{\Theta}}
\end{eqnarray}
with
\begin{eqnarray}
\tilde{R}&=&-\Delta\mathcal{Q}+\mathcal{L}_{E}E^{2}+\epsilon \Delta r^{2}-\mathcal{L}_{\phi}L_{\phi}^{2}-\mathcal{L}_{\psi}L_{\psi}^{2}+4abmL_{\phi}L_{\psi}
\nonumber\\
&&-4 a m (b^{2}+r^{2})EL_{\phi}-4bm(a^{2}+r^{2})EL_{\psi}\\
\tilde{\Theta}&=&\mathcal{Q}-L_{\phi}\csc^{2}\theta-L_{\psi}\sec^{2}\theta+(E^{2}+\epsilon)(\rho^{2}-r^{2}),
\end{eqnarray}
and
\begin{eqnarray}
\mathcal{L}_{E}&=&r^{2}\Delta+2m(a^{2}+r^{2})(b^{2}+r^{2})\\
\mathcal{L}_{\phi}&=&b^4-a^2 b^2-2 b^2 m-a^2 r^2+b^2 r^2\\
\mathcal{L}_{\psi}&=&a^4-a^2 b^2-2 a^2 m+a^2 r^2-b^2 r^2
\end{eqnarray}
For this case the celestial coordinates read
\begin{eqnarray}
\alpha&=&-(\xi_{1}\csc\theta+\xi_{2}\sec\theta)\\
\beta&=&\pm\sqrt{\eta+a^{2}\cos^{2}\theta+b^{2}\sin^{2}\theta-\xi_{1}^{2}\csc^{2}\theta-\xi_{2}^{2}\sec^{2}\theta}.
\end{eqnarray}

In what follows we shall explore the black holes shadow associated to the CCLP solutions for both $a=b$ and $a\ne b$ in the non--extremal and extremal cases. Given that the mass and the charge can be expressed in terms of the parameters $\{a,b,g\}$ the strategy we shall follow here is to set some values for $m$ and $q$ in the non--extremal cases and set the values given by the constraints (\ref{q}) and (\ref{gm}) for the extremal solution. 

\subsection{Case I: $\theta=\pi/2$; $a=b$}
\begin{figure*}[hbt!]
\centering
\includegraphics[width=0.3\textwidth]{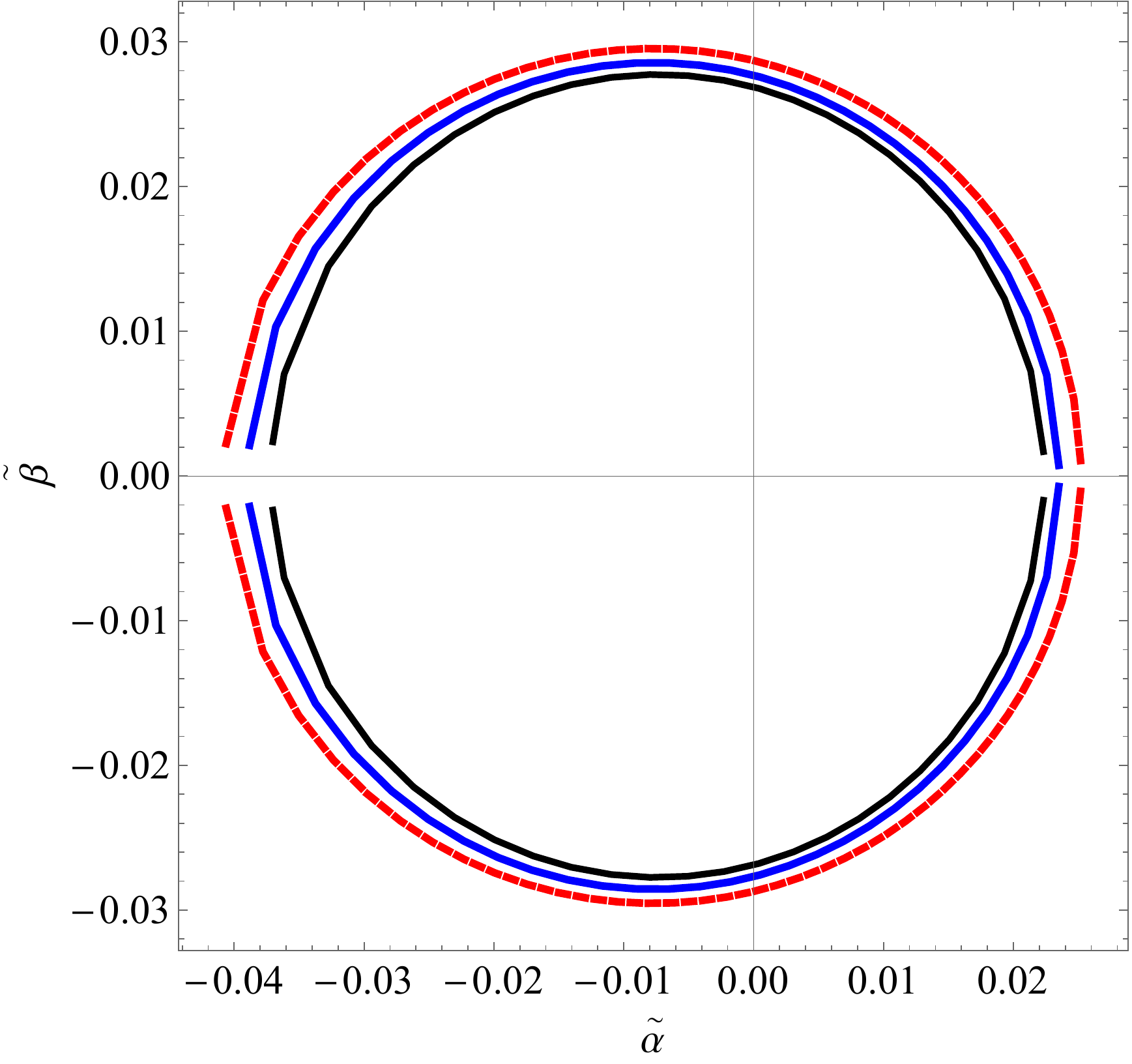}  \
\includegraphics[width=0.3\textwidth]{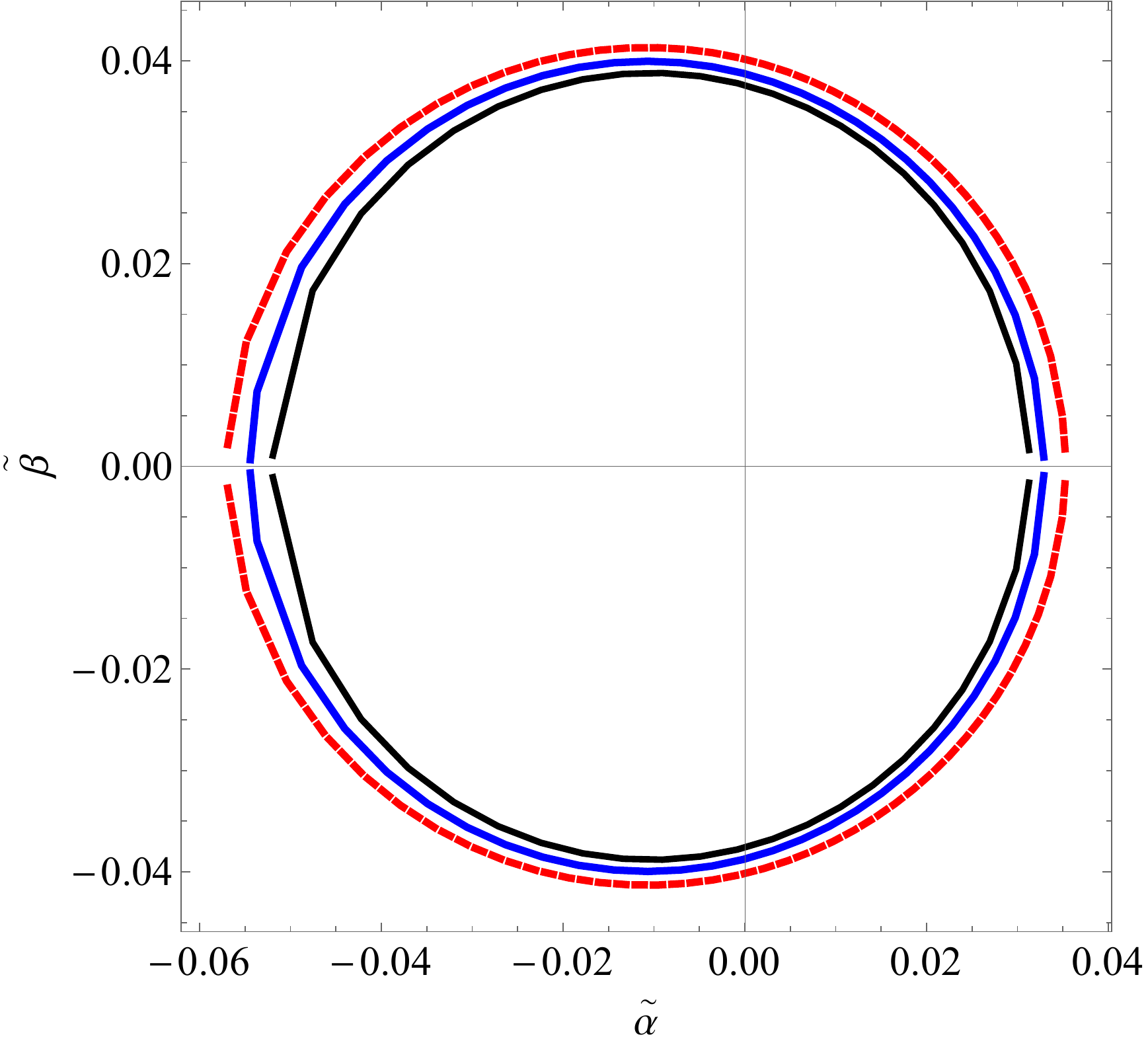}  \
\includegraphics[width=0.3\textwidth]{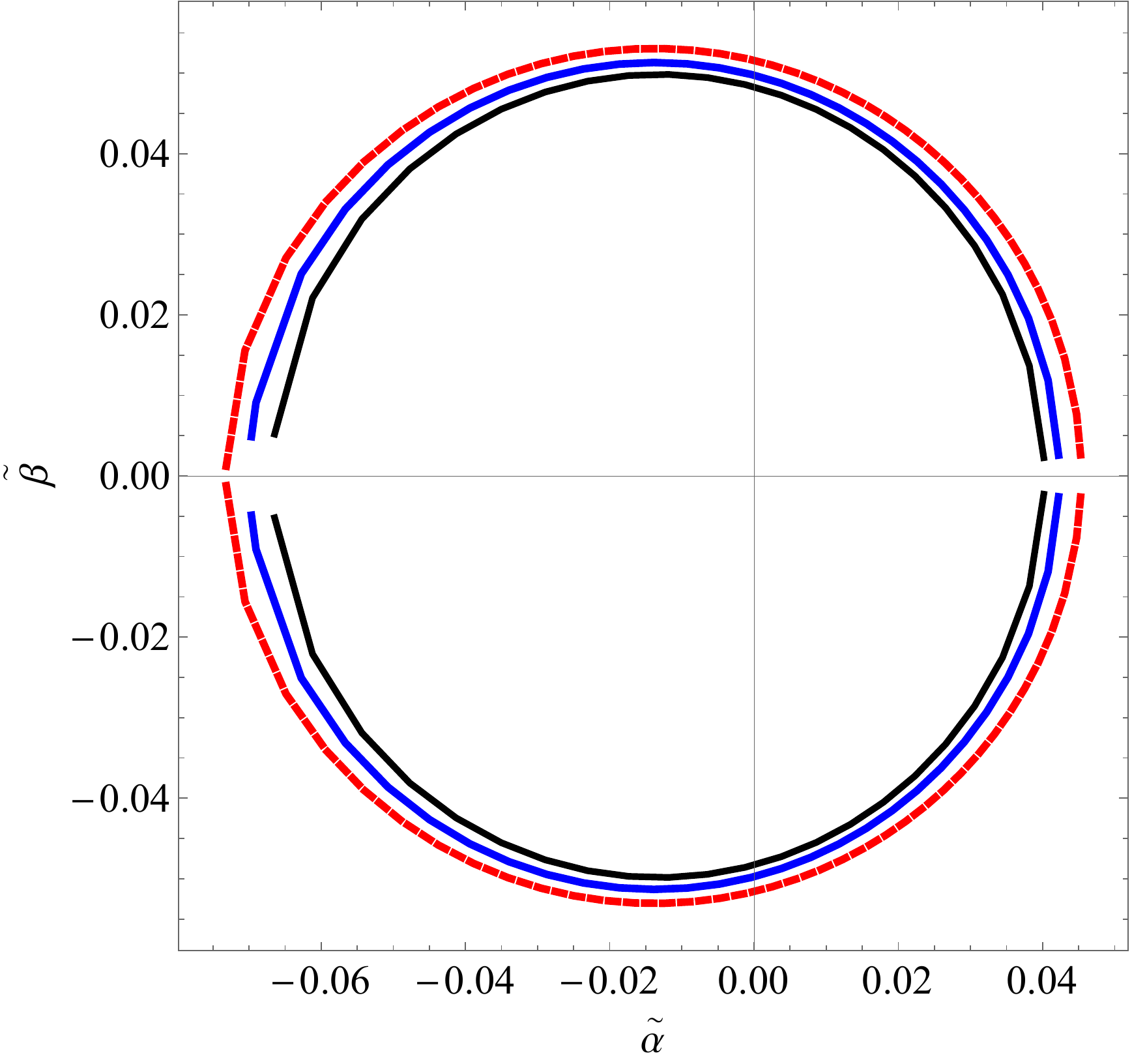} \
\caption{\label{figcaso1ne}BH shadow for the non extremal CCLP BH for
 $a=0.4$ (black), $a=0.45$ (blue) and $a=0.5$ (red) for $m=0.1$, $q=6$ and
 $g=0.05$ (left), $g=0.07$ (middle), and $g=0.09$ (right) }
\end{figure*}

\begin{figure*}[h!]
\centering
\includegraphics[width=0.3\textwidth]{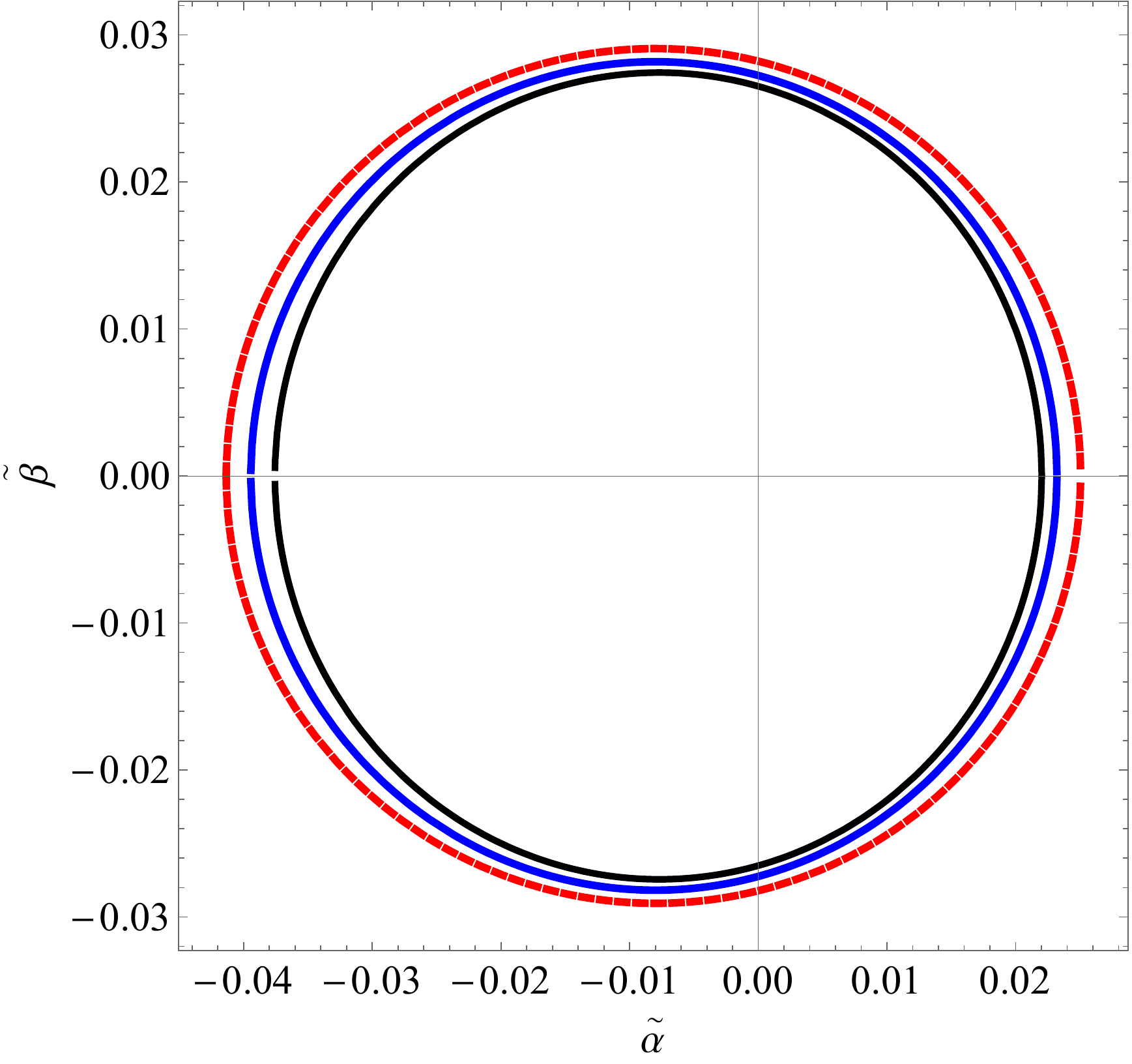}  \
\includegraphics[width=0.3\textwidth]{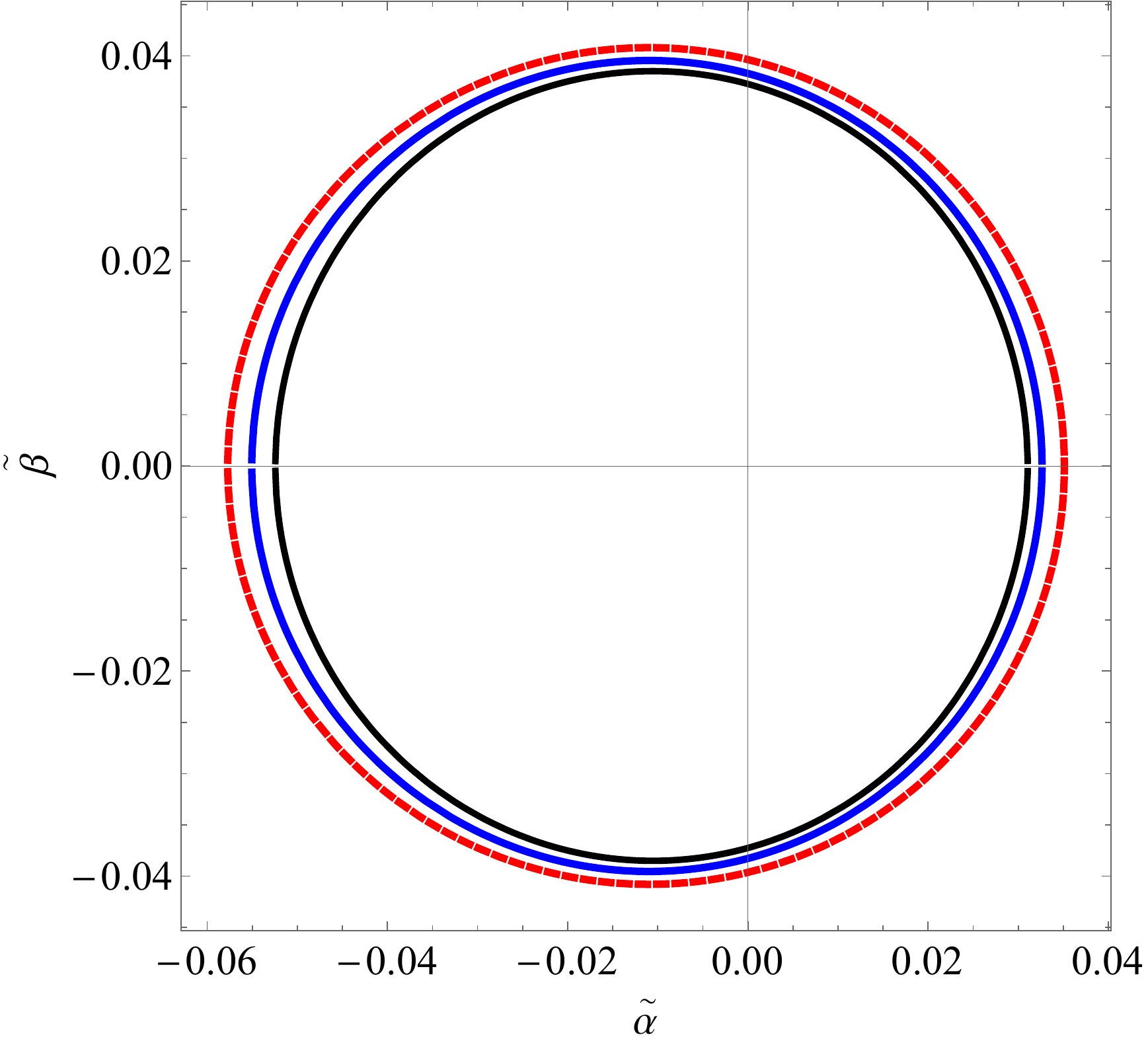}  \
\includegraphics[width=0.3\textwidth]{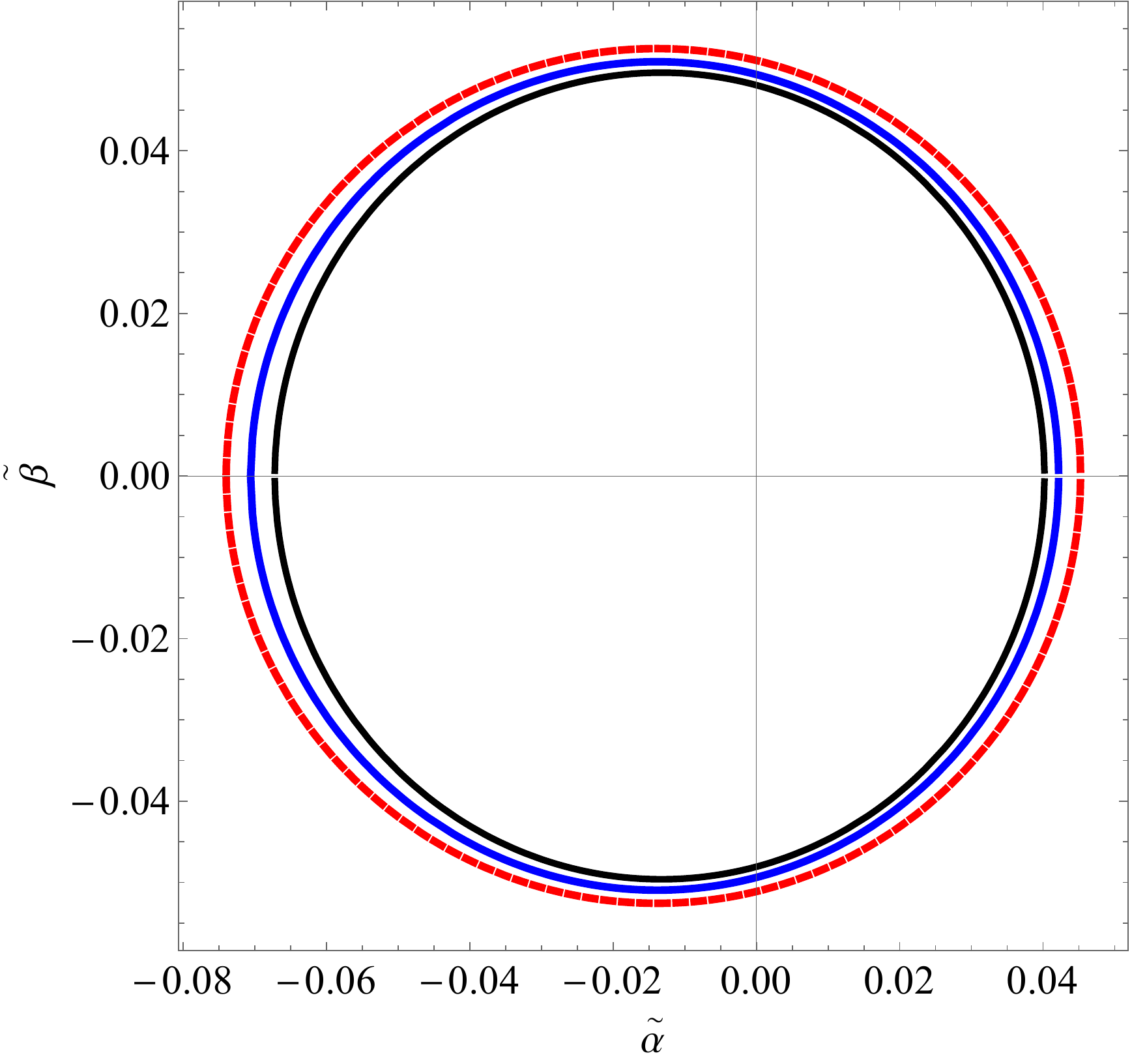} \
\caption{\label{figcaso1e}BH shadow for the extremal CCLP BH for
 $a=0.4$ (black), $a=0.45$ (blue) and $a=0.5$ (red) and
 $g=0.05$ (left), $g=0.07$ (middle), and $g=0.09$ (right) }
\end{figure*}

In fig. \ref{figcaso1ne} we show the shadow of the non--extremal CCLP BH for $a=b$. 
Note that for each panel shown in Fig. \ref{figcaso1ne} the size of the shadow increases as $a$ does. Furthermore, for the same $a$, the area of the shadow increases as $g$ increases. A similar behaviour is observed in the extremal case shown in fig. \ref{figcaso1e} 

\subsection{Case II: $\theta=\pi/2$; $a\ne b$}
\begin{figure*}[hbt!]
\centering
\includegraphics[width=0.3\textwidth]{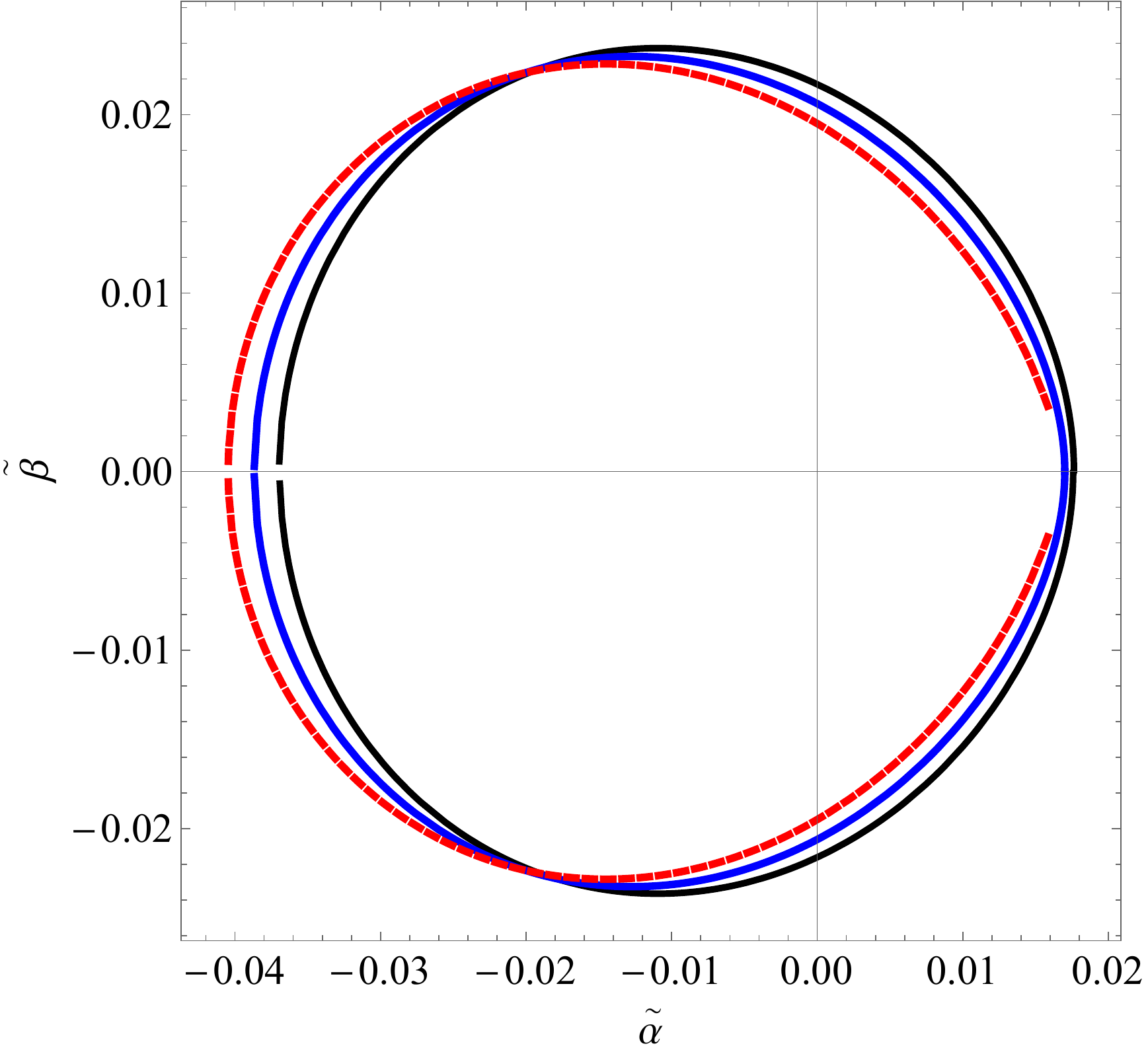}  \
\includegraphics[width=0.3\textwidth]{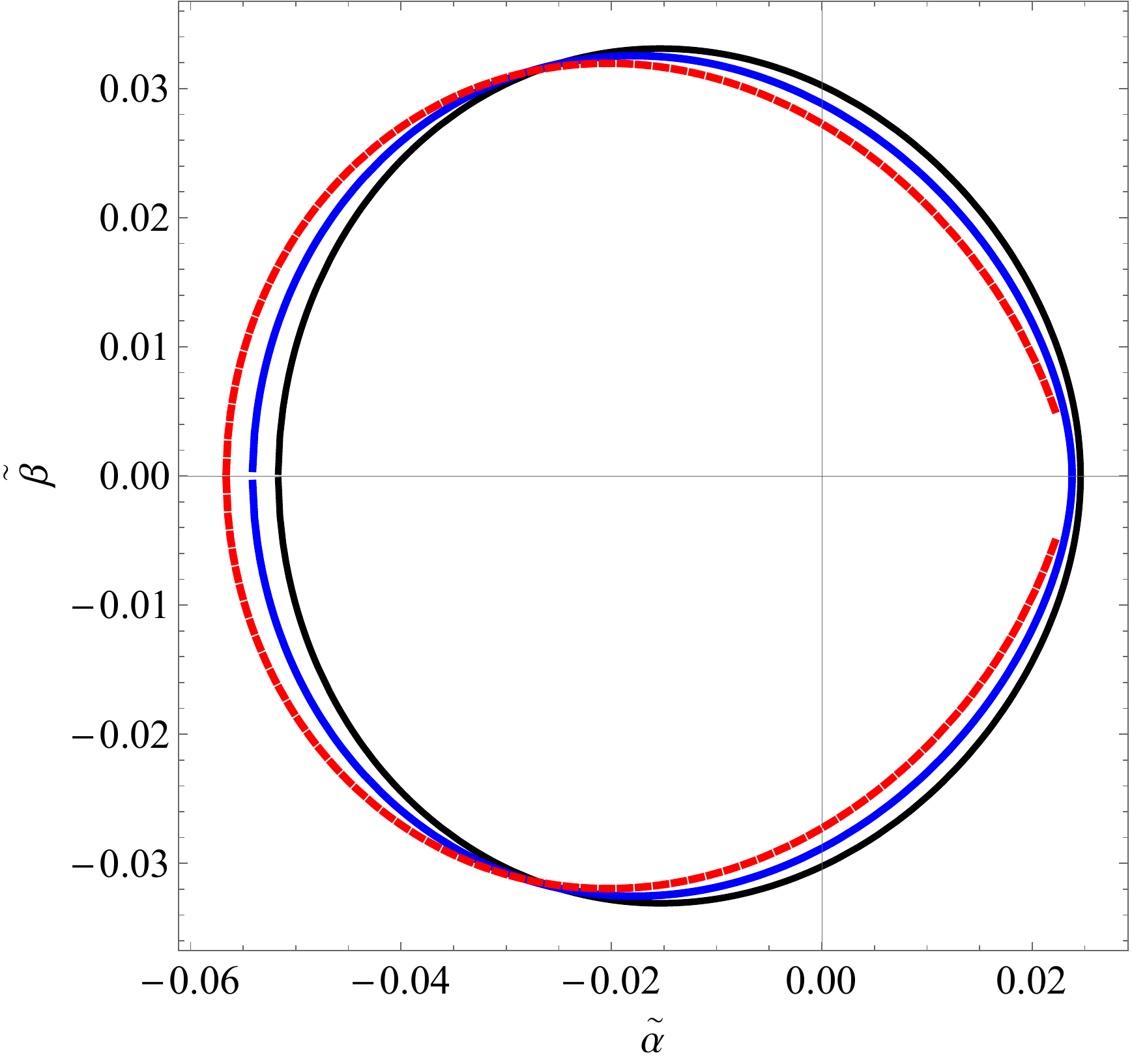}  \
\includegraphics[width=0.3\textwidth]{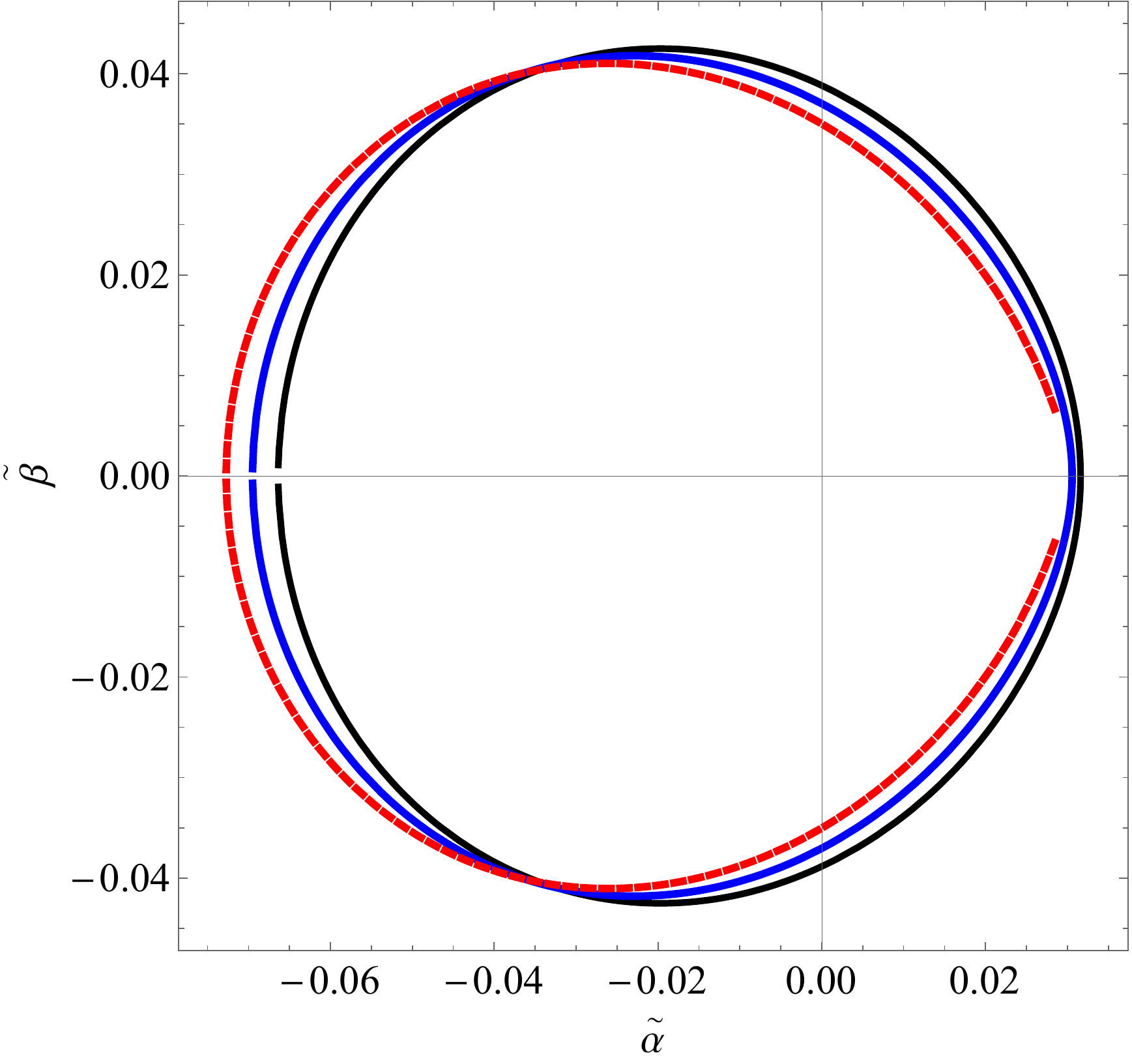} \
\caption{\label{figcaso2ne}BH shadow for the--non extremal CCLP BH for
 $a=0.4$ (black), $a=0.45$ (blue) and $a=0.5$ (red) for $m=0.1$, $q=6$, $b=0.1$ and
 $g=0.05$ (left), $g=0.07$ (middle), and $g=0.09$ (right) }
\end{figure*}

In fig. \ref{figcaso2ne} we show the shadow of the non--extremal CCLP BH for $a\ne b$.
Note that, in contrast to the previous case the shadow maintain the same area but the silhouette appears distorted. Besides, as $a$ increases the shadow shift to left.

\begin{figure*}[hbt!]
\centering
\includegraphics[width=0.3\textwidth]{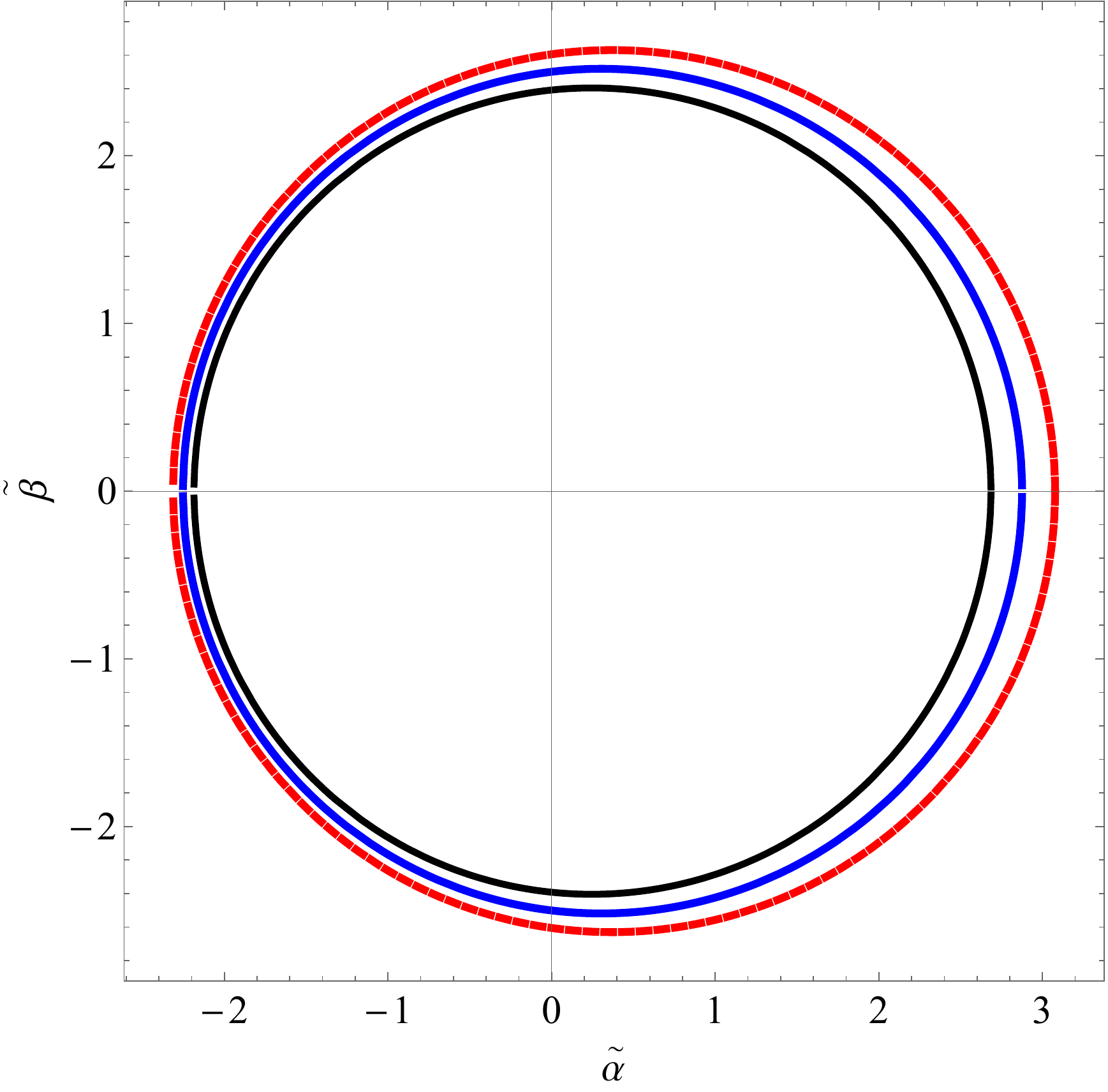}  \
\includegraphics[width=0.3\textwidth]{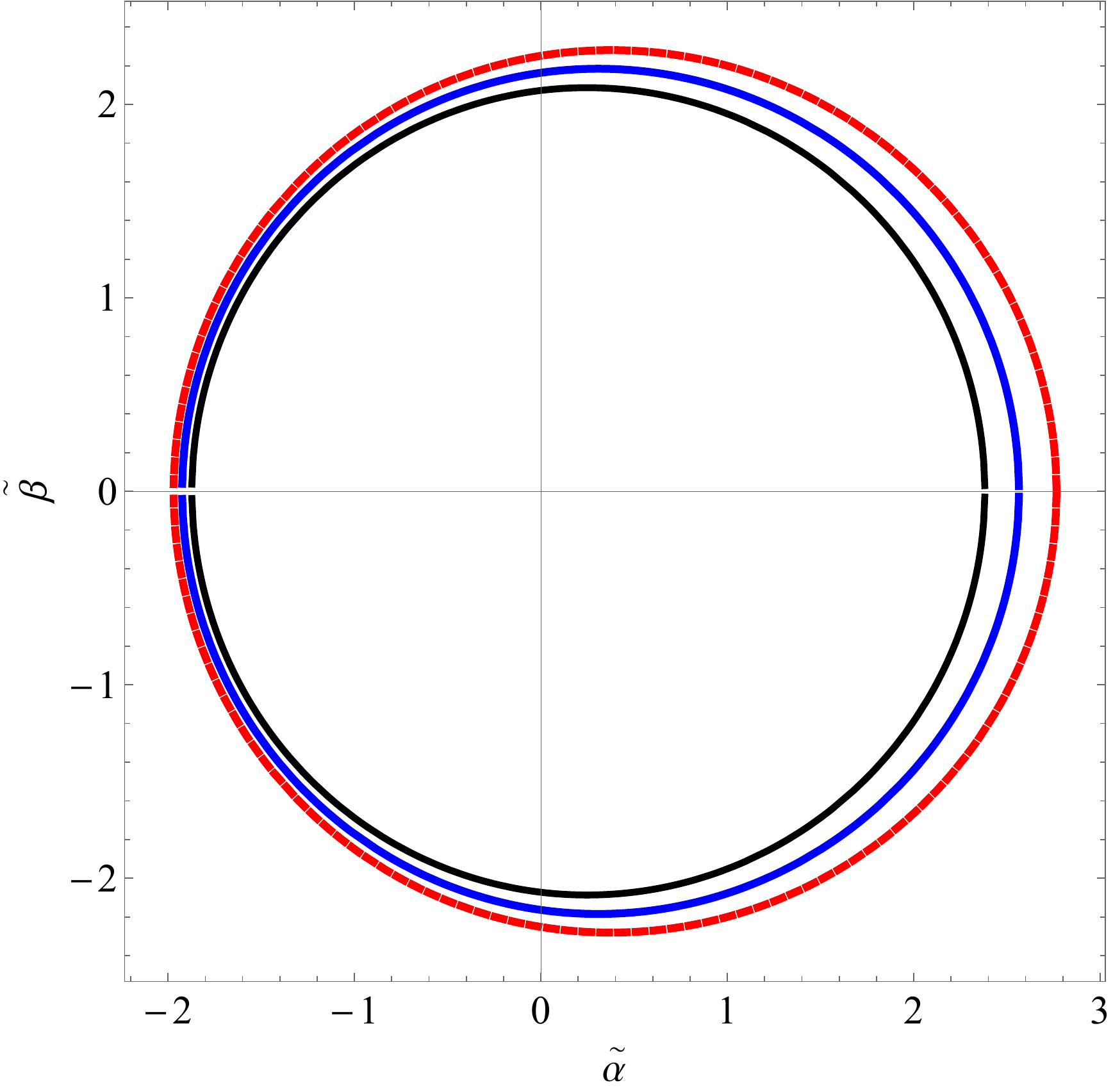}  \
\includegraphics[width=0.3\textwidth]{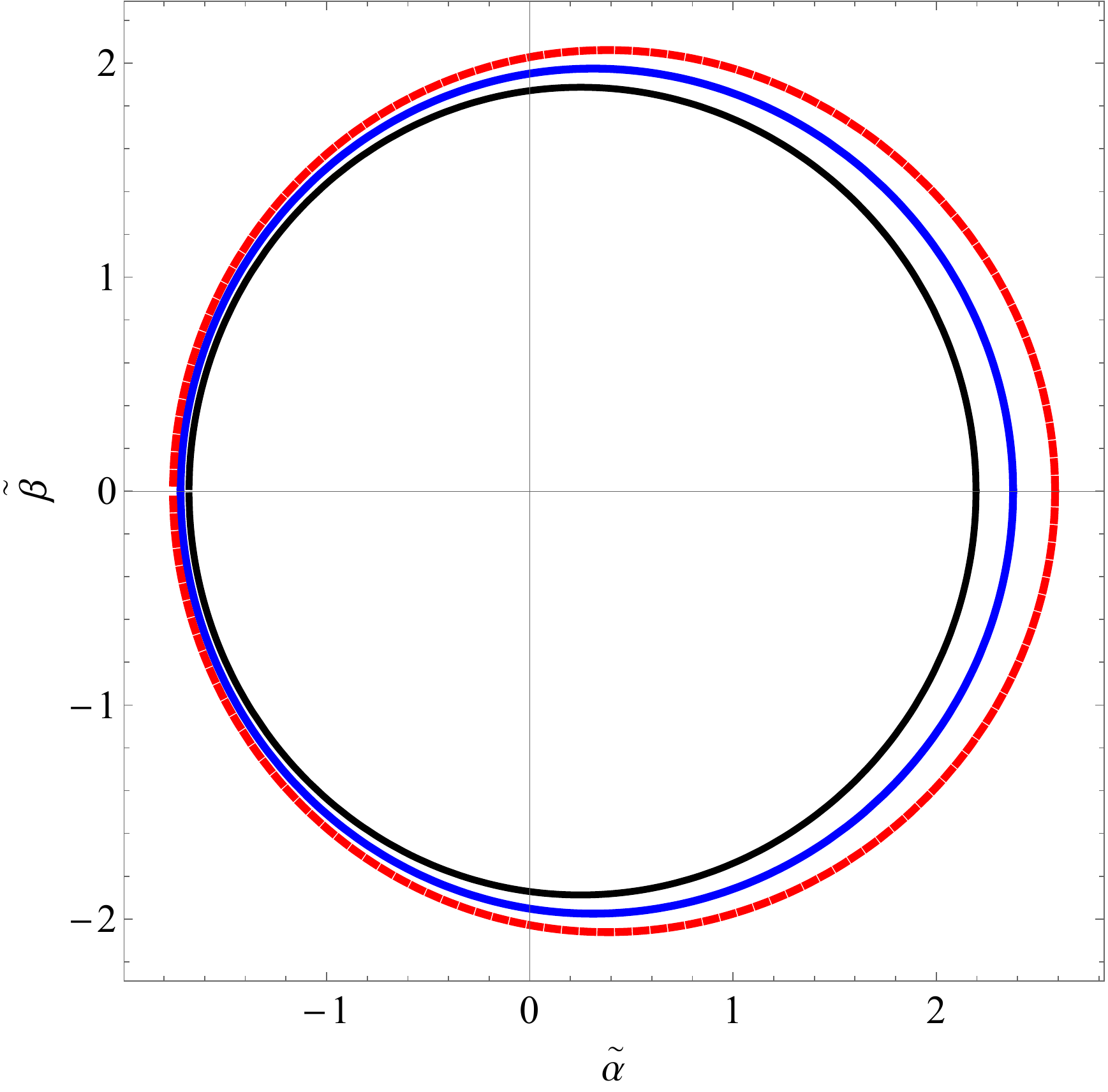} \
\caption{\label{figcaso2e}BH shadow for the extremal CCLP BH for
 $a=0.4$ (black), $a=0.45$ (blue) and $a=0.5$ (red) for $b=0.1$ and
 $g=0.05$ (left), $g=0.07$ (middle), and $g=0.09$ (right) }
\end{figure*}

In fig. \ref{figcaso2e} we show the shadow of the extremal CCLP BH.
Note that the shadow increases as
$a$ and $g$ increases in accordance to the case $a=b$. However, all the silhouettes appears to have the same value at $(-\tilde{\alpha},0)$ but the point $(\tilde{\alpha},0)$ moves to the right as $a$ increases. Besides, although the parameters involved are the same, the size of the shadow is appreciable more greater than in the previous cases. Indeed, the scale in Fig. (\ref{figcaso2e}) is at leas two orders of magnitude greater than in the rest of the plots.

\section{Conclusions}
Shadow cast by black holes in several different contexts has been a very active research field over the last years, and especially after the first image of a supermassive BH at the center of M87 two years ago by the Event Horizon Telescope. Moreover, theories with negative cosmological constant can be embedded in a supersymmetric setting obtaining gauged supergravity theories in various dimensions. Motivated by this fact, in this work we performed a detailed analysis of the geodesics around the Chong--Cveti\v{c}--L\"u--Pope
general non-extremal rotating BH in minimal five-dimensional gauged supergravity reported in \cite{39}. In particular we followed the  Hamilton-Jacobi formalism to obtain the geodesic equations of the solution for both massive and massless particles. Then, we analysed the effective radial potential and imposed the constraint to unstable orbits which allowed us to obtain the parametrical expressions for the impact parameters. As far as we know, this is the first time this procedure is presented in the literature. Besides, we deduced the coordinates of the celestial plane and compare this result with the obtained in the Myers-Perry case. To study the silhouette cast by the Chong--Cveti\v{c}--L\"u--Pope black hole, we considered equatorial geodesics defined by the vanishing of one of the angular momentum for both the non--extremal and extremal solutions. All the figures were plotted using the same set of parameters for a direct comparison, and as a result, we obtained that although the shape of the shadow appeared distorted in 
in some cases, their area was of the same order except for the extremal black hole with different angular parameters which size is around two orders greater that the others studied here.

\section*{Acknowlegements}
The author A. R. acknowledges DI-VRIEA for financial support through Proyecto 
Postdoctorado 2019 VRIEA-PUCV. The author G. P. thanks the Funda\c{c}ao para a Ciencia e Tecnologia (FCT),
Portugal, for the financial support to the Center for Astrophysics and Gravitation-CENTRA, Instituto Superior Tecnico, Universidade de Lisboa, through the Project No. UIDB/00099/2020 an to the Project No.  PTDC/FIS-AST/28920/2017.
P. B. is funded by the Beatriz Galindo contract BEAGAL 18/00207 (Spain).


\end{document}